\documentclass[useAMS,usenatbib]{mn2e}
\usepackage{graphicx}

\newcommand{\HII}{H\,{\sc ii}}
\newcommand{\NII}{[N\,{\sc ii}]}

\newcommand{\OII}{[O\,{\sc ii}]}

\newcommand{\Ha}{H$\alpha$}

\newcommand{\Ms}{$M_{\star}$}

\usepackage{savesym}
\usepackage{amsmath}
\savesymbol{iint}
\usepackage{txfonts}
\restoresymbol{TXF}{iint}

\bibpunct{(}{)}{;}{a}{}{,}

\title[GAMA: mass, metallicity, and SFR relationships]{Galaxy And Mass Assembly (GAMA): A deeper view of the  mass, metallicity, and SFR relationships}
\author[M. A. Lara-L\'opez \it {et. \,al}]{M. A. Lara-L\'opez$^{1}$\thanks{E-mail:
mlopez@aao.gov.au}\thanks{ARC Super Science Fellow}, A. M. Hopkins$^{1}$, A. R. L\'opez-S\'anchez$^{1,2}$, S. Brough$^{1}$
\newauthor
M.\,L.\,P.\,Gunawardhana$^{3,1}$, M. Colless$^{1}$, A. S. G. Robotham$^{4,5}$, A. E. Bauer$^{1}$
\newauthor
J. Bland-Hawthorn$^{3}$, M. Cluver$^{1}$, S. Driver$^{4,5}$, C. Foster$^{6}$, L. S. Kelvin$^{4,5,7}$, J. Liske$^{8}$
\newauthor
J. Loveday$^{9}$, M. S.  Owers$^{1}$,  T. J. Ponman$^{10}$, R. G. Sharp$^{11}$, O. Steele$^{12}$,  E. N. Taylor$^{3,13}$
\newauthor
D. Thomas$^{12}$\\
$^{1}$Australian Astronomical Observatory, PO Box 915, North Ryde, NSW 1670, Australia\\
$^{2}$Department of Physics and Astronomy, Macquarie University, NSW 2109, Australia.\\
$^{3}$Sydney Institute for Astronomy (SIfA), School of Physics, University of Sydney, NSW 2006, Australia\\
$^{4}$School of Physics \& Astronomy, University of St Andrews, North Haugh, St Andrews, KY16 9SS, UK\\
$^{5}$International Centre for Radio Astronomy Research, The University of Western Australia, 35 Stirling Highway, Crawley, WA 6009, Australia\\
$^{6}$European Southern Observatory, Alonso de Cordova 3107, Vitacura, Santiago, Chile\\
$^{7}$ Institut f\"{u}r Astro-und Teilchenphysik, Universit\"{a}t Innsbruck, Technikerstra\ss e 25, 6020 Innsbruck, Austria \\
$^{8}$European Southern Observatory, Karls-Schwarzschild-Str. 2, 85748 Garching, Germany\\ 
$^{9}$Astronomy Centre, University of Sussex, Falmer, Brighton BN1 9QH, UK\\
$^{10}$School of Physics and Astronomy, University of Birmingham, Edgbaston, Birmingham B15 2TT, UK\\
$^{11}$Research School of Astronomy \& Astrophysics, Australian National University, Cotter Road, Weston Creek, ACT 2611, Australia\\
$^{12}$Institute of Cosmology and Gravitation, University of Portsmouth, Dennis Sciama Building, Burnaby Road, Portsmouth PO1 3FX, UK\\
$^{13}$School of Physics, The University of Melbourne, Parkville, VIC 3010, Australia\\}

\begin{document}

\date{Accepted  . Received   ; in original form    }

\pagerange{\pageref{firstpage}--\pageref{lastpage}} \pubyear{2002}

\maketitle

\label{firstpage}

\begin{abstract}

A full appreciation of the role played by gas metallicity ($Z$), star-formation rate (SFR), and stellar mass (\Ms) is fundamental to understanding how galaxies form and evolve. The connections between these three parameters at different redshifts  significantly affect galaxy evolution, and thus  provide important constraints for galaxy evolution models. Using data from the Sloan Digital Sky Survey--Data Release 7 (SDSS--DR7) and the Galaxy and Mass Assembly (GAMA) surveys we  study the relationships and dependencies between SFR, $Z$, and \Ms, as well as the Fundamental Plane for star-forming galaxies. We combine both surveys using volume-limited samples up to a redshift of $z \approx$ 0.36. The GAMA and SDSS surveys complement each other when analyzing the relationships between SFR, \Ms\ and $Z$. We present evidence for SFR and metallicity evolution to  $z \sim$0.2. We study the dependencies between SFR, \Ms\, $Z$, and specific star-formation rate (SSFR) on the \Ms--$Z$, \Ms--{\it SFR}, \Ms--{\it SSFR}, $Z$--{\it SFR}, and $Z$--{\it SSFR} relations, finding strong correlations between all. Based on those dependencies, we propose a simple model that allows us to explain the different behaviour observed between low and high mass galaxies. Finally, our analysis  allows us to confirm the existence of a Fundamental Plane, for which \Ms=$f(Z, {\rm SFR})$ in star-forming galaxies.

\end{abstract}


\section{Introduction}

% HABLAR GENERALMENTE DE LA MASA, SFR Y METALICIDAD

% Mergers, interactions, dense environments, gas inflow and outflows can trigger or suppress the SFR, increase the metallicity, 

The star formation history and  chemical enrichment are two of the main  parameters that drive the evolution of galaxies. A detailed appreciation of those properties spanning several cosmological epochs will provide stringent constraints on how galaxies form and evolve.

The stellar mass (\Ms),  star-formation rate (SFR), and gas metallicity  ($Z$) have been related in the past through the well known mass-metallicity  (\Ms$-Z$) \citep[e.g.][]{Lequeux79, Tremonti04}, and mass-SFR (\Ms$-$SFR) relationships \citep[e.g.][]{Brinchmann08, Noeske07a}, Also, it has been shown that there is no strong correlation directly between the metallicity and SFR in galaxies \citep[e.g.][]{Lara10b,LopezSanchez10, Yates11}.

The \Ms$-Z$ relation quantifies how the mass and metallicity of galaxies are related, with massive galaxies showing higher metallicities than less massive galaxies. Since metallicity is a tracer of the fraction of baryonic mass that has been converted into stars and is sensitive to the metal losses due to stellar winds, supernovae, and active galactic nuclei (AGN) feedback, the \Ms$-Z$ relation provides essential insight into galaxy formation and evolution. The \Ms$-Z$ relation has been extensively studied in the local universe \citep[e.g.][among others]{Lequeux79, Tremonti04,Kewley08}. With the advent of integral field spectroscopy (IFS) surveys, new results on the origin of the \Ms$-Z$ relation have been explored. Recently, \citet[][]{Rosales12} demonstrate the existence of a local \Ms$-Z$ relation using 2572 spatially resolved \HII\ regions in 38 galaxies. Furthermore, \citet[][]{Sanchez13} obtain the same local \Ms$-Z$ relation using CALIFA IFS data.

Metallicity has been shown to evolve to lower values even out to relatively low redshifts of $z \approx$ 0.4 \citep[e.g.][]{Lara09a, Lara09b, Pilyugin11}. As redshift increases, the metallicity evolution is stronger.  All studies of the  \Ms$-Z$ relation in high-redshift ($z\sim$ 0.7) galaxies have shown an evolution in metallicity with respect to local galaxies  \citep{Savaglio05, Maier05, Hammer05, Liang06,Rodrigues08}. At redshift $z\sim$2.2, \citet{Erb06} found that galaxies have a lower metallicity by $\sim$ 0.3 dex, while at redshift $z\sim$3.5, \citet{Maiolino08}  reported a strong metallicity evolution, suggesting that this redshift corresponds to an epoch of major star-formation activity. The \Ms$-Z$ relation has  also been studied for different morphological  galaxy types.  In particular, \citet{Calura09} found that, at any redshift, elliptical galaxies have the highest stellar masses and the highest stellar metallicities, whereas the least massive and chemically unevolved objects are the irregular galaxies, see also \citet{Pilyugin13} for a observational result.

There are two main ways to explain the origin of the \Ms$-Z$ relation. The first  is attributed to metal and baryon loss due to gas outflow, where low--mass galaxies eject large amounts of metal--enriched gas by supernovae winds before high metallicities are reached, while massive galaxies have deeper gravitational potentials that retain their gas, thus reaching higher metallicities \citep{Larson74,Dekel86,MacLow99,Maier04,Tremonti04,DeLucia00,Kobayashi07,Finlator08}. As pointed out in the high--resolution simulations of \citet{Brooks07}, supernovae feedback plays a crucial role in lowering the star formation efficiency in low--mass galaxies. Without energy injection from supernovae to regulate the star formation, gas that remains in galaxies rapidly cools, forms stars, and increases the metallicity too early, producing a \Ms$-Z$ relation too flat compared to observations.

A second scenario to explain the \Ms$-Z$ relation is related to the well-known effect of downsizing \citep[e.g.][]{Cowie96,Gavazzi96}, in which lower mass galaxies form their stars later and on longer time-scales than more massive systems, implying low star formation efficiencies in low--mass galaxies \citep[e.g.][]{Efstathiou00,Brooks07,Mouhcine08,Tassis08,Scannapieco08,Ellison08}. Therefore, low--mass galaxies are expected to show lower metallicities.  Supporting this scenario, \citet{Calura09} reproduced the \Ms$-Z$ relation with chemical evolution models for ellipticals, spirals and irregular galaxies, by means of an increasing efficiency of star formation with mass in galaxies of all morphological types, without the need for outflows favoring the loss of metals in the less massive galaxies. Also, \citet{Vale09} modeled the time evolution of stellar metallicity using a closed-box chemical evolution picture, suggesting that the \Ms$-Z$ relation for galaxies in the mass range from $10^{9.8}$ to $10^{11.65}\,M_{\odot}$ is mainly driven by the star formation history and not by inflows or outflows.

% AQUI HABLAR SOBRE LA SFR

The SFR is a key parameter to understand the stellar evolution. 
% The development of direct SFR diagnostics includes the integrated emission-line fluxes, near ultraviolet continuum fluxes, and infrared continuum fluxes, see \citet{Kennicutt98} for a review. Alternatively, it is possible to estimate the SFR from the soft X-ray luminosity, which is comparable to that determined from the {H$\alpha$} luminosity \citep{RosaGon09, Rolivos09}. 
The hydrogen Balmer lines  (mainly, the H$\alpha$ line) are   one of the most reliable tracers of star formation \citep[e.g.][]{Moustakas06}, since the Balmer emission-line luminosity scales directly with the total ionizing flux of the embedded stars  in {H\,\textsc{ii}} regions and star-forming galaxies.  It is important however, to take into account corrections for stellar absorption and obscuration to obtain SFRs in agreement with those derived using other wavelengths \citep[e.g.][]{RosaGon02, Charlot02, Dopita02, Hopkins03}. 

% Among the SFR calibrations with the {H$\alpha$} line, we have \citet{Kennicutt98}.  In parallel, other diagnostics have been developed using the oxygen doublet [{O\,\textsc{ii}}] $\lambda$3726, 3729 for the redshift range $z \sim 0.4-1.5$ \citep[e.g.][]{Gallagher89, Kennicutt98, RosaGon02, Kewley04}. Moreover, this diagnostic is usefull when the {H$\alpha$} line is not easily observable at higher redshifts ($z \gtrsim 0.4$ in the optical). However, the [{O\,\textsc{ii}}] doublet presents problems in reddening and abundance dependence \citep[e.g.][]{Jansen01, Charlot02}.

A strong dependence of  the SFR with the stellar mass, as well as its evolution with redshift has been found, with the bulk of star formation occurring first in massive galaxies, and later in less massive systems \citep{Guzman97, Brinchmann00, Juneau05, Bauer05, Bell05,PerezGon05, Feulner05, Papovich06, Caputi06, Reddy06, Erb06, Noeske07a, Buat08}. In the local universe, several studies have illustrated a relationship between the SFR and stellar mass, identifying two populations: galaxies on a star-forming sequence, and $``$quenched" galaxies, with little or no detectable star formation \citep{Brinchmann04, Salim05}. At higher redshift, \citet{Noeske07a} showed the existence of a $``$main sequence" (MS) for SF galaxies in the \Ms--{\it SFR} relation over the redshift range $0.2 < z < 1.1$.  \citet{Noeske07a} show that the slope of the MS remains constant to $z>1$, while the MS as a whole moves to higher SFR as $z$ increases.

% ESTO ES PARA HABLAR SOBRE LOS PLANOS

The existence of fundamental planes (FP) is a natural result of scaling relationships between important astrophysical properties.
The first study proposing a relationship between these three fundamental quantities was by \citet{Ellison08} who found a SFR dependence on the \Ms$-Z$ relation for star forming (SF) galaxies using data from the SDSS. A FP was found by \citet{Lara10a} in a three dimensional study of the \Ms\, gas metallicity, and SFR of SF galaxies using data from the SDSS-DR7.  \citet{Lara10a}  showed that the \Ms$-Z$, and \Ms--{\it SFR} relationships are particular cases of a more general relationship, a FP. This combination reduces the scatter significantly compared to any other pair of correlations. In a parallel study, \citet{Mannucci10} found similar fundamental 3D relationships,  fitting a surface to the stellar mass, SFR, and gas metallicity.   \citet{Mannucci10} provided an expression, referred to as the Fundamental Metallicity Relation (FMR), in which $Z$ is expressed as a combination of \Ms\ and SFR.  In this regard,  \citet{Yates11}  used star-forming galaxies from the SDSS--DR7  to study the dependencies of the different combinations of SFR, metallicity and stellar mass.  These authors found  that, although having high dispersion, there is a dependence of \Ms\ on the $Z$--{\it SFR} relation,  and they obtained similar  dependencies using models.  \citet{Yates11} also found that the fit given by \citet{Mannucci10} does not significantly reduce the dispersion of the metallicity compared to the \Ms$-Z$ relation. 

Just in the past few years, many studies have attempted to quantify the distribution of galaxies in this three dimensional space \citep{Magrini12, Peebles13}. Through testing a variety of different methodologies,  \citet{Lara12} conclude that a planar distribution is sufficient to account for 98$\%$ of the variance in the \Ms$-Z-${\it SFR} space.

% In this paper, we aim to clearify, using a highly statistical sample, the metallicity evolution of galaxies up to redshift $\sim$0.35, which up to know has been seen only in small samples \citep{Lara09a, Lara09b, Pilyugin11}. 

This paper is structured as follows. In \S\,\ref{SampleSelection} we detail the data used for this study. In \S\,\ref{EvolutionS} we  analyze the evolution of the SFR, SSFR and metallicity for our sample of galaxies. In \S\,\ref{FP} we investigate the Fundamental Plane for GAMA and SDSS galaxies. In \S\,\ref{dependencies}  we present the different dependencies of SFR, metallicity and \Ms. Finally, in \S\,\ref{Conclusions} we give a summary and conclusions. Throughout we assume $H_0=70\,$km\,s$^{-1}$\,Mpc$^{-1}$, $\Omega_M=0.3$, $\Omega_{\Lambda}=0.7$.

%__________________________________________________________________

\section[]{Sample selection}\label{SampleSelection}

\begin{figure*}
\includegraphics[scale=0.48]{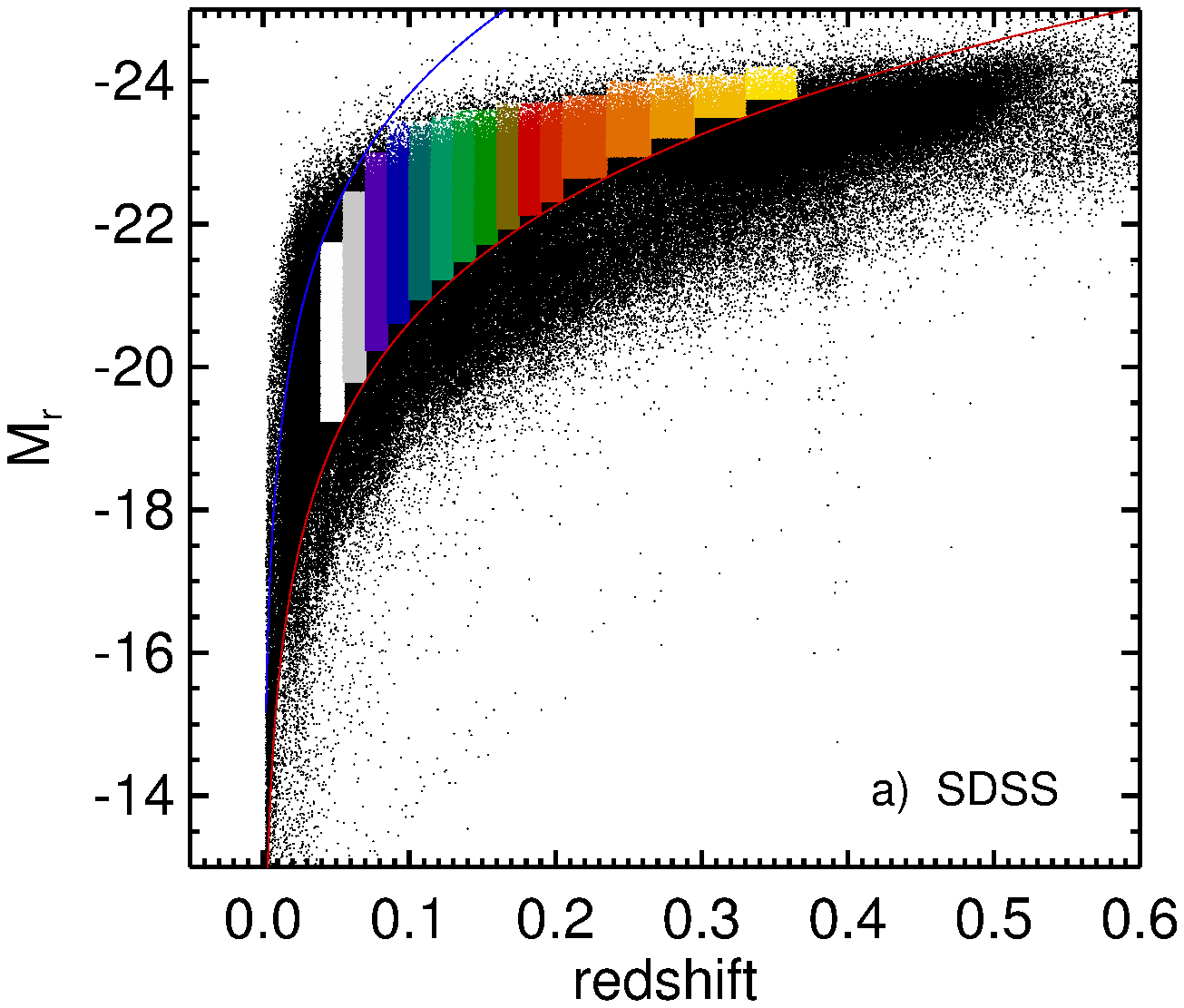}
\includegraphics[scale=0.48]{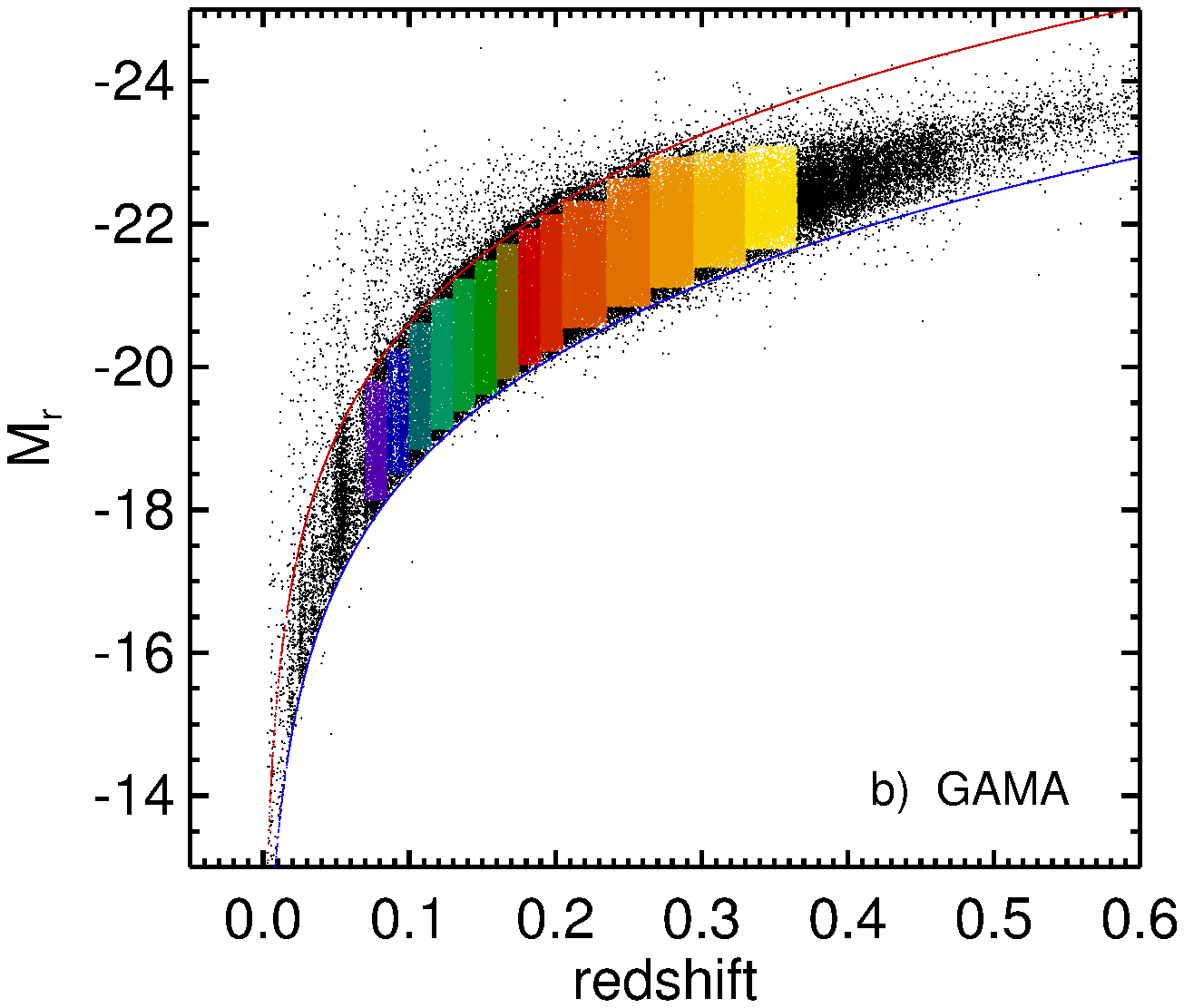}
\caption{Redshift vs. Petrosian $r$-band absolute magnitude ($M_r$) for the (a) SDSS and (b) GAMA surveys. The red line is the same in both plots, and  corresponds to the bright/faint apparent $r$-magnitude ($m_r$) limit of 17.77 for GAMA/SDSS. The blue  lines correspond, respectively, to the  SDSS $m_r$ limit of 14.5, and to the  GAMA $m_r$ limit of 19.8. The colour boxes correspond to each of the volume limited samples, where the same colours represent the same redshift bins, but different magnitudes for each survey. Note that the white and grey boxes in the SDSS sample correspond to redshifts 0.04 $<z<$ 0.07 that do not appear in the GAMA sample.}
\label{RedshiftVsMr}
\end{figure*}

We consider  emission-line galaxies from two large surveys, the ``Galaxy and Mass Assembly" (GAMA) survey \citep{Driver11}, and the  ``Sloan Digital Sky Survey--Data Release 7" (SDSS--DR7) \citep{Abaza09,Adelman07}.

GAMA is a spectroscopic survey based on data taken with the 3.9m Anglo-Australian Telescope (AAT) using the 2dF fibre feed and AAOmega multi-object spectrograph \citep{Sharp06}. The spectra were taken with  2 arcsec diameter fibre, a spectral coverage from 3700 to 8900 {\AA}, and spectral resolution of 3.2 {\AA}. In this study we use the GAMA phase--I survey, which covers three fields of 48 $\rm{deg}^2$, with Petrosian magnitude limits of $m_{r}<19.8$ mag in one field, and $m_{r}<19.4$ mag in the other two. The GAMA data used in this paper include spectra of $\sim$ 140,000 galaxies. Our main galaxy sample for GAMA is composed of galaxies in the range $17.77 < m_{r} < 19.8$ and redshifts up to $z \approx 0.36$.

Emission lines for the GAMA survey are measured in two ways. As a first approach, we fit Gaussians to a selection of common emission lines at appropiate observed wavelengths, given the measured redshift of each object. The local continuum spanning each fitting region is approximated with a linear fit. A second approach uses the Gas AND Absorption Line Fitting algorithm \citep[GANDALF,][]{Sarzi06} to measure emission lines for the GAMA galaxies. GANDALF is a simultaneous emission and absorption line fitting algorithm designed to separate the relative contribution of the stellar continuum and of nebular emission in the spectra of galaxies, while measuring the gas emission and kinematics. GANDALF measures and corrects for dust attenuation and stellar absorption in the emission lines. The final set of measurements for both methods includes the flux, equivalent width, and signal--to--noise ratio for each emission line, among other results. Both measurements show a good agreement between the independent approaches \citep{Hopkins13}.

For SFR estimations the first approach is used. The {H$\alpha$} line is measured directly from the flux-calibrated spectra, corrected for dust using the Balmer Decrements and for stellar absorption as detailed by \citet{Brough11} and \citet{Mad11}. For metallicity measurements, we use the GANDALF catalogue. The GANDALF measurements account for the stellar absorption in the Balmer lines through the SED fitting. This doesn't improve the SFR estimates, due to the approach taken in making obscuration corrections within GANDALF, but it does provide the most robust estimate of the adjacent line ratios used in the metallicity estimates.

Data from the SDSS were taken with a 2.5 m telescope located at Apache Point Observatory \citep{Gunn06}. The SDSS spectra were obtained using 3 arcsec diameter fibres, covering a wavelength range of 3800-9200 {\AA}, a spectral resolution $\lambda$/$\Delta\lambda$ $\sim$1800-2200, and a wavelength coverage from 3800-9200\, \AA. The SDSS--DR7 spectroscopy database contains spectra for $\sim$ $\rm 1.6 x {10}^6$ objects, including 929,555 galaxies over $\sim$ 9380 $\rm{deg}^2$. Further technical details can be found in \citet{Stoughton02}.

We used the emission-line analysis of SDSS-DR7 galaxy spectra from the Max-Planck-Institute for Astrophysics--John Hopkins University (MPA-JHU)  database\footnote{http://www.mpa-garching.mpg.de/SDSS/}. Apparent and absolute Petrosian magnitudes were taken from the STARLIGHT database\footnote{http://www.starlight.ufsc.br} \citep{Cid05, Cid07, Mateus06, Asari07}. From the full dataset, we only consider objects classified as galaxies in the $``$main galaxy sample$"$ \citep{Strauss02} with apparent Petrosian $r$ magnitude in the range $14.5 < m_{r} < 17.77$.

\begin{figure*}
\includegraphics[scale=0.38]{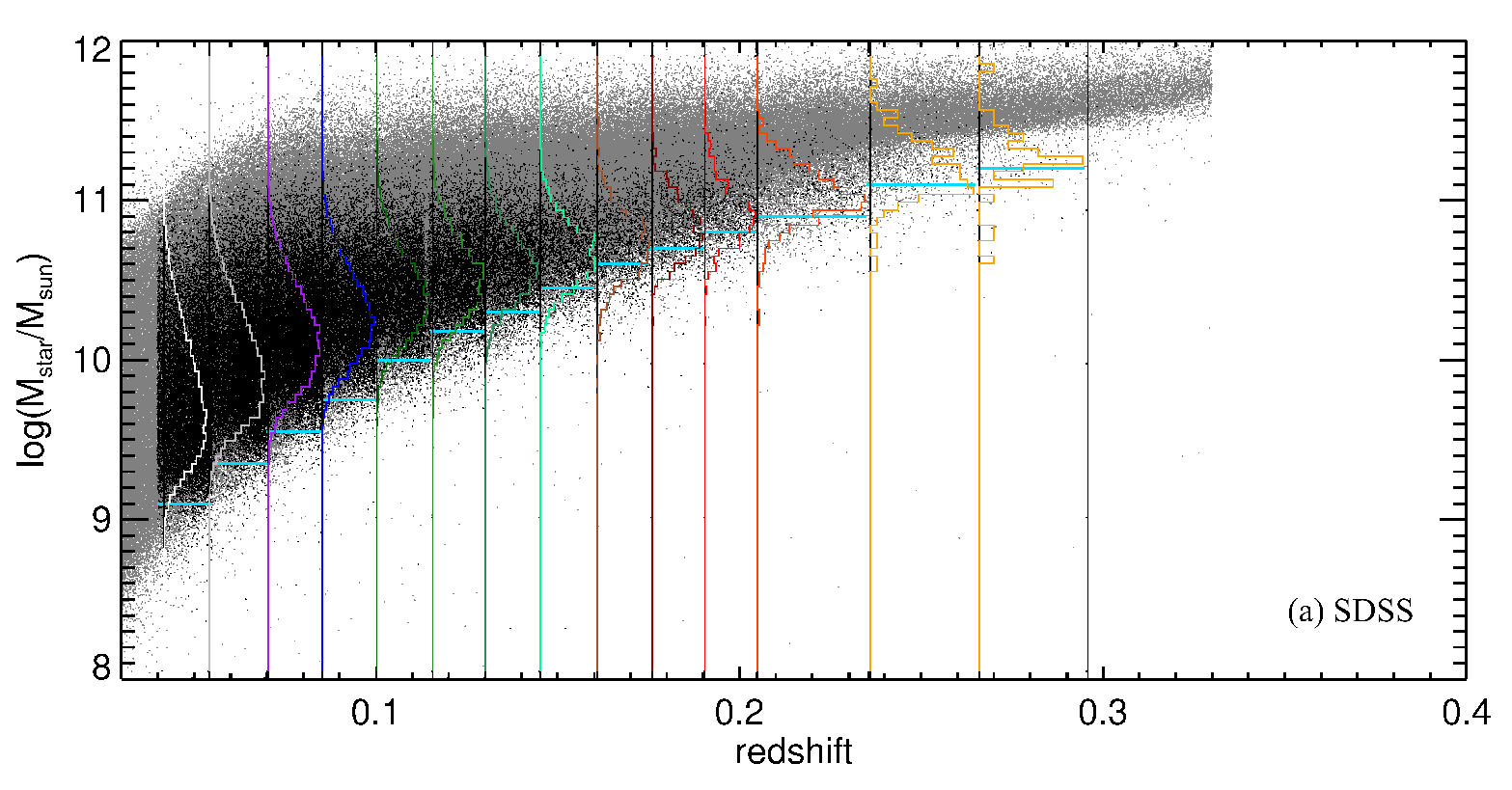}
\includegraphics[scale=0.38]{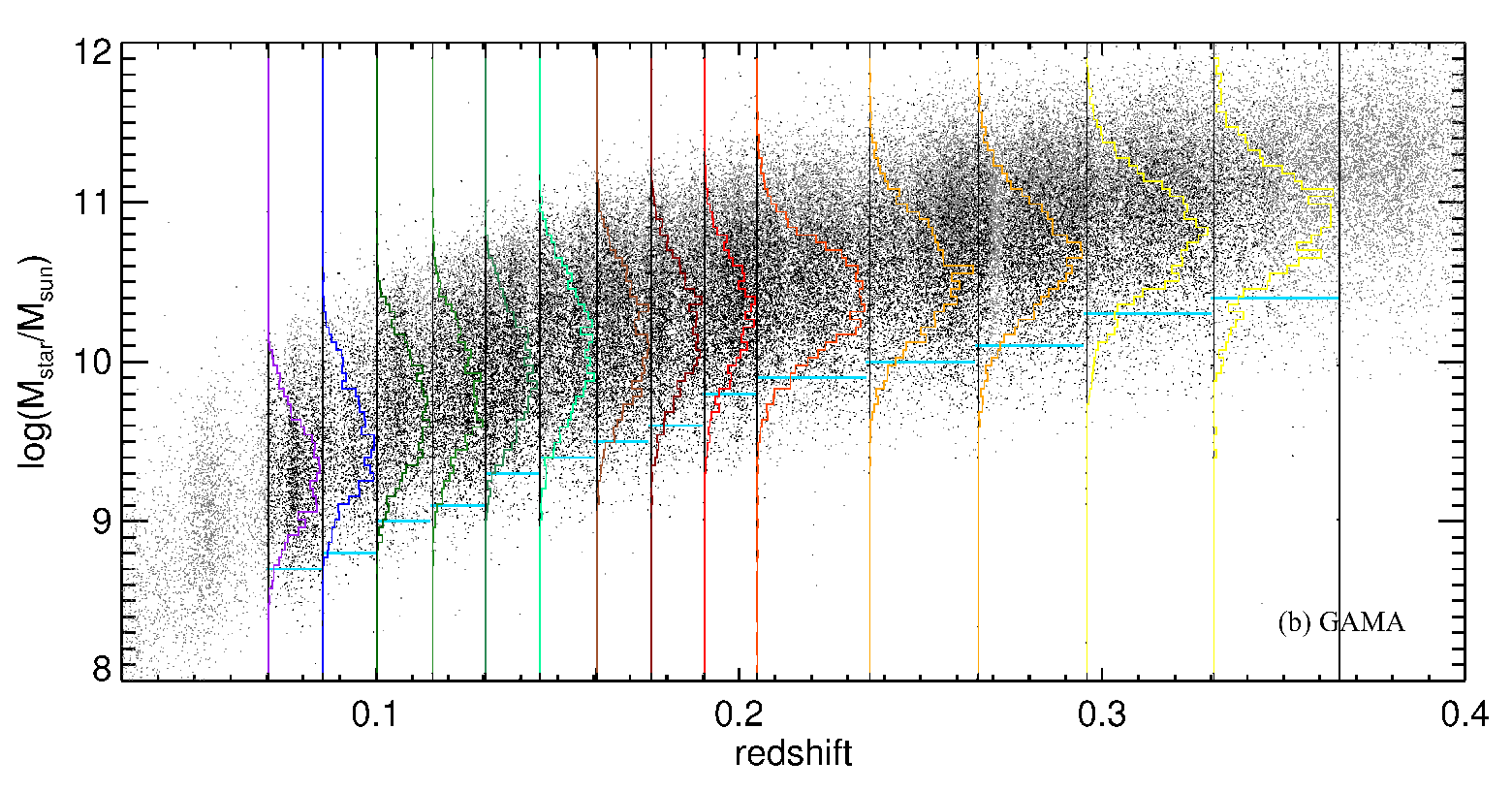}
\caption{Redshift vs. stellar mass for (a) SDSS  and (b) GAMA. In both panels, grey dots show the full sample, while black dots show only the volume limited samples of Fig. \ref{RedshiftVsMr}. The vertical histograms in each plot correspond to galaxies in each volume limited sample color coded as in previous plot. Horizontal light blue lines correspond to the mass limit in each redshift bin.}
\label{RedshiftVsMasa}
\end{figure*}

\begin{table*}
\begin{center}
\begin{tabular}{ccc||cc||cc||}
\hline
\hline
\multicolumn{3}{c} {}&\multicolumn{2}{c} {SDSS}&\multicolumn{2}{c}{GAMA}\\\cline{1-7}
\hline
Sample  & z$_{\rm min}$ & z$_{\rm max}$ &M$_r$ (upper) & M$_r$ (lower) & M$_r$ (upper) &  M$_r$ (lower) \\\hline

1	&	0.04		&	0.055	&	-19.248	&	-21.734	&	--	&	--	\\\
2	&	0.055	&	0.07		&	 -19.795	&	 -22.448	&	--	&	--	\\\
3	&	0.070	&	0.085	&	-20.239	&	-22.995	&	-18.139	&	 -19.795	\\\
4	&	0.085	&	0.1		&	-20.615	&	-23.439	&	-18.515	&	-20.239	\\\
5	&	0.1		&	0.115	&	-20.950	&	-23.439	&	-18.850	&	 -20.615	\\\
6	&	0.115	&	0.130	&	-21.227	&	-23.5	&	-19.127	&	-20.950	\\\
7	&	0.130	&	0.145	&	-21.485	&	-23.6	&	-19.385	&	 -21.227	\\\
8	&	0.145	&	0.160	&	-21.719	&	-23.6	&	 -19.619	&	21.485	\\\
9	&	0.160	&	0.175	&	-21.933	&	-23.7	&	-19.833	&	-21.719	\\\
10	&	0.175	&	0.190	&	-22.131	&	-23.7	&	 -20.031	&	-21.933	\\\
11	&	0.190	&	0.205	&	-22.316	&	-23.7	&	-20.216	&	-22.131	\\\
12	&	0.205	&	0.235	&	-22.649	&	-23.8	&	-20.549	&	-22.316	\\\
13	&	0.235	&	0.265	&	-22.946	&	-24.0	&	-20.846	&	-22.649	\\\
14	&	 0.265	&	0.295	&	-23.213	&	-24.1	&	-21.113	&	-22.946	\\\
15	&	 0.295	&	0.330	&	-23.494	&	-24.1	&	-21.394	&	-23.0	\\\
16	&	0.330	&	0.365	&	 -23.750	&	-24.2	&	-21.650	&	-23.1	\\\hline
\noalign{\smallskip}
\end{tabular}
\normalsize
\rm
\end{center}
\caption{Redshifts and $r$-band absolute Petrosian magnitude limits for the volume limited samples of the SDSS and GAMA surveys shown in Fig. \ref{RedshiftVsMr}.}
\label{VolSamples}
\end{table*}

According to \citet{Kewley05},  a minimum of 20$\%$ of the galaxy light inside the optical fibre is needed  to avoid any possible bias due to the fibre diameter.  Hence, we imposed a lower redshift limit of  $z$=0.04 and $z$=0.07 for data  from the SDSS and GAMA samples, respectively. The redshift difference is due to the different fibre diameters used in each survey. With these redshift limits any metallicity biases due to the fibre aperture sampling only the central regions of a galaxy should be minimised.

% {\bf **** Fibras unicas tampoco dan bien SFR, para eso se hace SAMI en particular. Discutir, referencias ****}

From the GAMA and SDSS samples described above, we construct volume limited samples by selecting narrow redshift bins of equal absolute Petrosian $r$-band magnitudes, as shown in Fig.\ref{RedshiftVsMr}a,b and Table \ref{VolSamples}. To construct the volume limited samples we used the full data samples of both surveys. The red line in both panels of Fig.\ref{VolSamples} shows the apparent Petrosian $r$-band bright/faint limit for GAMA/SDSS, respectively, corresponding to $m_{r}$=17.77. The blue line for the GAMA sample, Fig.\ref{RedshiftVsMr}b, corresponds to the  limit in apparent Petrosian magnitude $m_{r}$=19.8, while for the SDSS sample, Fig.\ref{RedshiftVsMr}a, the blue line corresponds to the  limit of  $m_{r}$=14.5. The limits on the stellar mass of the volume limited samples were determined as shown by the horizontal blue lines  in Fig. \ref{RedshiftVsMasa}.

\begin{figure}
\includegraphics[scale=0.6]{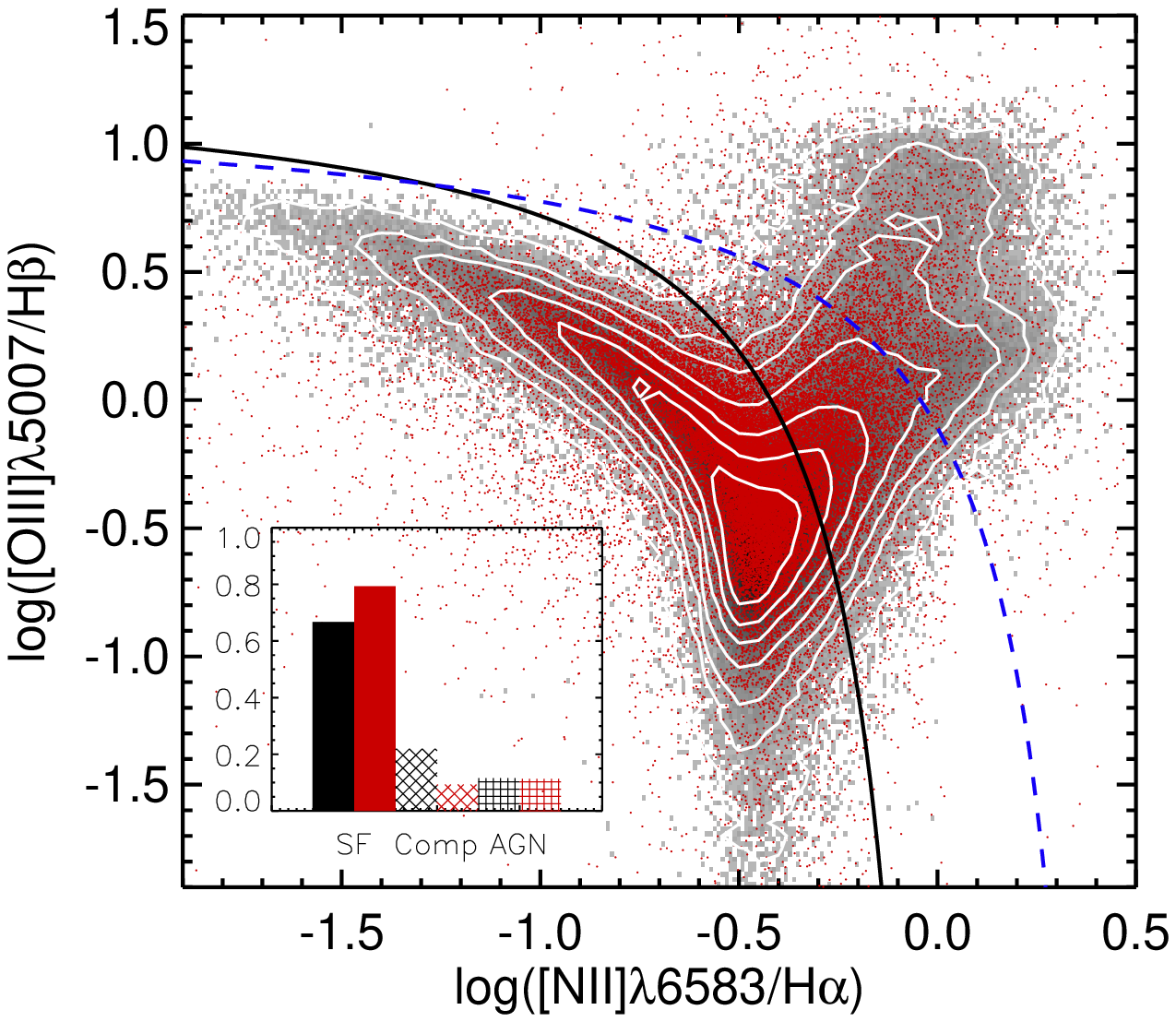}
\caption{BPT \citep{Baldwin81} diagram for the SDSS and GAMA samples. The grey density plot and white contours correspond to the SDSS sample, while the red dots correspond to the GAMA sample. The solid  line is the empirical relation provided by \citet{Kauf03a}, and galaxies below this line correspond to SF galaxies. The dashed line corresponds to the \citet{Kewley01} relation. Galaxies between the solid and dashed lines correspond to composite, while galaxies above the dashed line correspond to AGNs. The histogram shows the percentage of SF, composite and AGN galaxies for the SDSS (black) and the GAMA (red) samples.}
\label{BPT}
\end{figure}

For both galaxy surveys we select all galaxies for which data in each of the  H$\alpha$, H$\beta$, [{N\,\textsc{ii}}]~$\lambda$6584, and [{O\,\textsc{iii}}]~$\lambda$5007 emission lines are available. We obtain a total of 85,378 and 632,652  for the GAMA and SDSS surveys, respectively. Our next step is to select only SF galaxies  and not objects with some kind of nuclear activity (e.g. composite and AGN galaxies). To this end, we use the  BPT diagram \citep{Baldwin81},  that compares the [{O\,\textsc{iii}}]~$\lambda$5007/H$\beta$ and [{N\,\textsc{ii}}]~$\lambda$6584/H$\alpha$ ratios (Fig. \ref{BPT}). We also impose a signal--to--noise ratio (SNR) higher than 3$\sigma$ for {H$\alpha$},  {H$\beta$}, and \NII. This last criterion results in 455,872 objects for SDSS and 28,238 galaxies for GAMA.

From this  sample, we classify the objects into SF, composites and AGN galaxies following the criteria of \citet{Kauf03a} and \citet{Kewley01}.  The corresponding percentage of SF, composite, and AGN galaxies for GAMA is 79.3$\%$, 9.3$\%$, and 11.4$\%$,  respectively, while for SDSS it is 66.5$\%$, 21.9$\%$, 11.5$\%$, respectively.  Fig. \ref{BPT} shows the BPT diagram for both samples.  This figure also includes a histogram showing the percentage of SF, composites, and AGN galaxies for both surveys. The  GAMA survey has more SF galaxies than SDSS  because GAMA is  deeper  than the SDSS, and more sensitive to low-mass systems at low redshift, which are more dominated by star formation than AGN activity.

% **********SERSIC INDEX*****

\begin{figure}
\begin{center}
\includegraphics[scale=0.42]{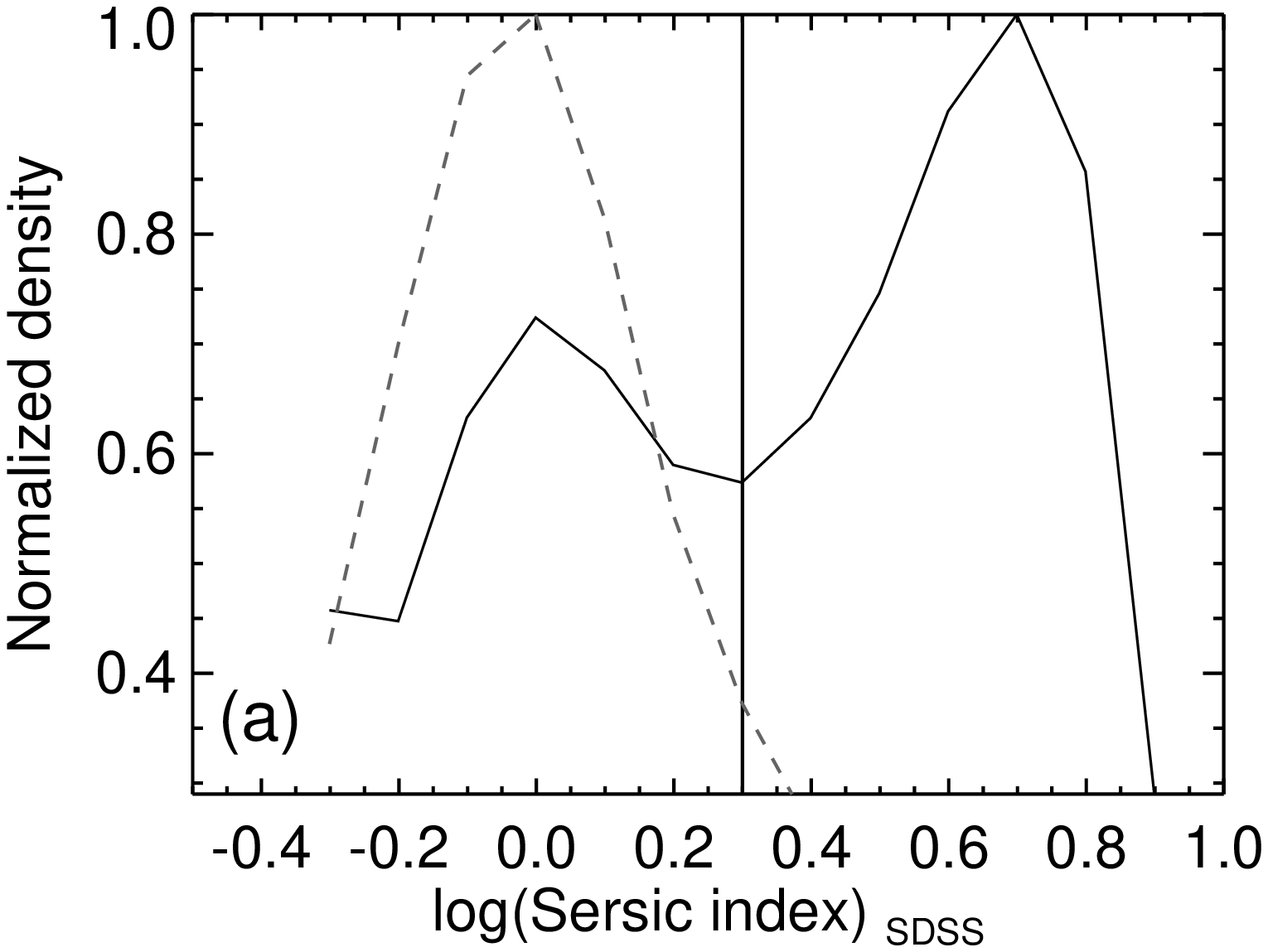}
\includegraphics[scale=0.42]{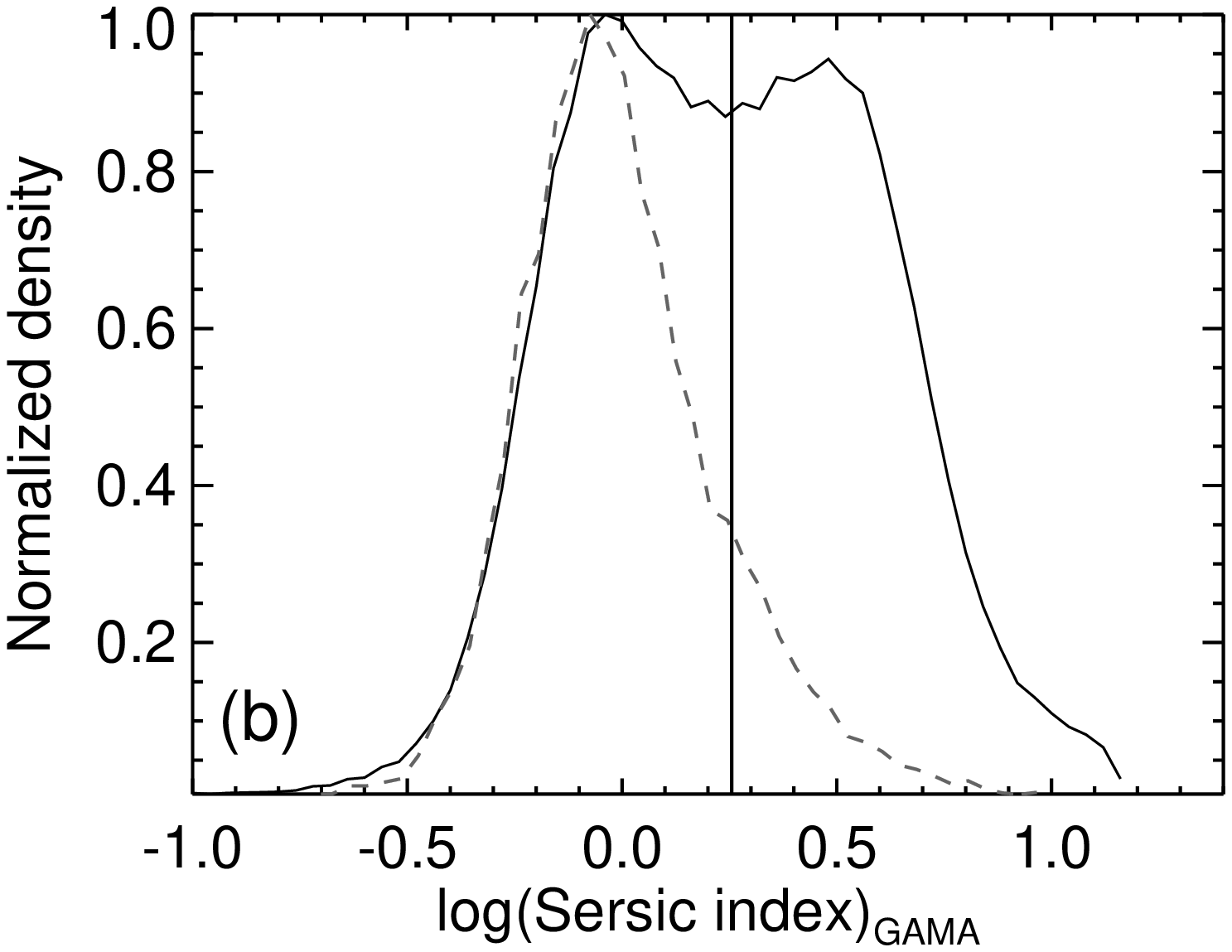}
\caption{Normalized histograms of the S\'ersic index for the whole galaxy sample of the (a) SDSS and (b) GAMA surveys. In both panels, the solid black line corresponds to the whole sample, while the dashed  line corresponds to the SF sample. The vertical solid line indicates the limit between late-- and early--type galaxies.}
\end{center}
\label{Sersix}
\end{figure}

\citet{Kelvin11} estimate the S\'ersic indexes ($n$) for galaxies in the GAMA survey via a detailed and independent modeling in the $ugriz$, $Y$, $J$,  $H$ $\&$ $K$ bands.  \citet{Kelvin11} report a bimodality, with two Gaussian-like distributions in most of the bands. For the $r$-band there is a rough separation at $n\approx$1.9 [$\log(n) \approx 0.278$] between early-- and late--type galaxies. Based on the S\'ersic index, 86$\%$ of the GAMA--SF sample used in  our study correspond to  late--type galaxies (Fig. \ref{Sersix}).

For the SDSS survey, global S\'ersic indexes (n$_g$) were estimated by \citet{Simard11} fitting simultaneous bulge+disk decompositions in $g$ and $r$ bands. These authors only provide a single S\'ersic index for both bands. A value of $n_g\sim$2 [$\log(n) \approx 0.3$] , gives a rough separation between the two distributions. According to this, 80$\%$ of our SDSS-SF sample correspond to late--type galaxies (Fig. \ref{Sersix}).

The distribution of the GAMA sample shows a similar Gaussian--bimodality between both populations of late-- and early--type galaxies. However, since the SDSS is a survey of brighter galaxies, it shows a stronger early--type population of galaxies. There is a small systematic difference between the S\'ersic indexes of both surveys due to the different estimation methods used. Nevertheless, both methods indicate that most of our SF sample correspond to late-type galaxies.

\subsection[]{SFR, Metallicity and Stellar mass estimates}

\subsubsection[]{The SDSS sample}

We use several methods to determine  the gas-phase metallicities and SFRs for the SF galaxies of the SDSS survey. Metallicities were estimated  using ({\it i})   the  empirical calibration  provided by  \citet{Pettini04}  between the oxygen abundance and the O3N2 index (which is defined below);
({\it ii}) the calibration given by \citet{Kewley02}, with the update of \citet{Kewley08}, which  is based on photo-ionization models  and considers the [{N\,\textsc{ii}}] / [{O\,\textsc{ii}}] ratio,  and ({\it iii}) the \citet{Tremonti04} metallicities, which  are also based on photo-ionization models and rely on Bayesian methods.

We compare these three methods and find that the  scatter in metallicity for the \Ms$-Z$ relation is minimised when \citet{Tremonti04} is used. Hereafter, we use those metallicities for the  data from the SDSS sample. \citet{Tremonti04} estimated metallicities statistically using Bayesian techniques based on simultaneous fits of  the most prominent emission lines ([{O\,\textsc{ii}}], {H$\beta$}, [{O\,\textsc{iii}}], {H$\alpha$}, [{N\,\textsc{ii}}], [{S\,\textsc{ii}}]), using a model designed for the interpretation of integrated galaxy spectra \citep{Charlot01}. Since the metallicities derived with this technique are discretely sampled, they exhibit small random offsets  \citep[for details see][]{Tremonti04}. %Any dependence of SFR on the estimated metallicity would be minor \citep{Tremonti04, Brinchmann08}.

We estimate SFRs using two different approaches.  First, we use the method described by \citet{Hopkins03},  which uses the equivalent widths (EW) of the {H$\alpha$} line. We correct the SFRs for stellar absorption, obscuration, and aperture effects (see \S\,\ref{SecGAMA} for details).  We also use \citet{Brinchmann04}, whose SFRs are based on Bayesian methods. A comparison between both SFRs shows a  tight correlation (see Appendix~A for details). While the SFRs derived using \citet{Hopkins03} are robust, for the SDSS sample we choose to use the total SFRs of \citet{Brinchmann04}.  As we show below, our results do not change significantly regardless of which SFR estimate we use. \citet{Brinchmann04} estimated SFRs modelling the emission lines in the galaxies following the  \citet{Charlot02} prescription, achieving a robust dust correction. The metallicity-dependence for the case B recombination in the {H$\alpha$}/{H$\beta$} ratio is also taken into account.

% Total stellar masses were estimated from fits to the photometry using the same modeling methodology as described in Kauffmann et al (2003), with only small differences with respect to previous data released.

Total stellar masses were estimated as in \citet{Kauf03a}, which relies on spectral indicators of the stellar age, and the fraction of stars formed in recent bursts.  These authors used the $z-$band magnitude to characterize the galaxy luminosity and constrain the star formation history using both the 4000\AA\ break, $D_n$(4000), and the stellar  Balmer absorption, H$\delta_A$.
The location of a galaxy in the $D_n$(4000)--H$\delta_A$ plane is insensitive to reddening and  depends weakly on metallicity. A \citet{Kroupa01} stellar initial mass function (IMF) was assumed.

\subsubsection[]{The GAMA sample}\label{SecGAMA}

From the GAMA SF sample described in $\S\,\ref{SampleSelection}$, we computed metallicities using the O3N2 parameter, which is defined as 

\begin{equation} \label{O3N2}
\rm O3N2 \equiv  log\left({\frac{[{\rm O\,\textsc{iii}}] \;  \lambda 5007/{\rm H}\beta}{[{\rm N\,\textsc{ii}}]  \; \lambda 6583/{\rm H}\alpha}}\right),
\end{equation} applying the calibration of  \citet{Pettini04}:

\begin{equation} \label{PP04}
\rm [ 12+log(O/H) ]_{\rm PP04} = 8.73 - 0.32 \times \rm O3N2
\end{equation}

However, it is well known \citep[e.g.][]{LSE10b,Moustakas+10} that there is an offset of $\sim$0.3 dex between the oxygen abundances derived using empirical calibrations such as this, which rely on direct estimations of the electron temperature of the ionized gas  \citep{Pettini04} and those derived using photoionization models \citep{Kewley02} or \citet{Tremonti04}. This issue has been recently reviewed  by \citet{LSD12}.
Hence, to work in the same system as the MPA-JHU Bayesian estimates, we compute metallicities using Eq.\ref{O3N2} and \ref{PP04} for the SDSS, and calibrate them to the \citet{Tremonti04} metallicities (T04), obtaining:

\begin{equation}
{\rm [  12+log(O/H) ] }_{\rm T04}=0.1026 + 1.0211 \times {\rm [ 12+log(O/H) ]}_{\rm PP04}
\end{equation} 
Following this equation, a galaxy with an oxygen abundance of 12+log(O/H)=9.00 derived following the \citet{Pettini04} calibration would correspond to a metallicity of 12+log(O/H)=9.29 from the  \citet{Tremonti04} method. See Appendix A for details of this calibration, and Appendix B for a comparison between GAMA and SDSS metallicities. For an analysis of metallicity errors and signal to noise effects for the GAMA sample see \citet{Foster12}.

SFRs were estimated using the prescription of \citet{Hopkins03} which relies on the EW$_{\rm H{\alpha}}$, and  corrected for obscuration, stellar absorption and fibre aperture as follows:

\begin{equation} \label{SFRsHop}
{\rm SFR}_{\rm Hopkins}=\frac{{L}_{{\rm H}\alpha}}{1.27 \times 10^{34}},
\end{equation} where  $L_{{\rm H}{\alpha}}$ corrected by stellar absorption and dust obscuration is given by

% { \small
 \begin{equation} \label{LumHop} \begin{split}
L_{\rm{H{\alpha}}} = & (EW_{\rm H{\alpha}}+ EW_c) {\times} 10^{-0.4(M_r-34.10)} {\times} \frac{3 {\times} 10^{18}}{[6564.61(1+z)]^2} \\
 &  \times \left(\frac{{ F({H{\alpha}})_{\rm obs}}/{ F({H{\beta}})}_{\rm obs}}{2.86}\right)^{2.36}
\end{split}
\end{equation} and $F$(${\rm H{\alpha}}$) and $F$(${\rm H{\beta}}$) are the observed emission line fluxes of H${\alpha}$ and H${\beta}$, respectively.  $EW_c$ corresponds to a fixed absorption stellar correction of 0.7 \citep{Mad11}, and $M_r$ is the absolute Petrosian magnitude \citep[for a detailed discussion see][]{Hopkins03}.
% , for details,  and 2.36 corresponds to the extinction corrected factor for the H$\alpha$ line using the  extinction curve of \citet{Cardelli89}, see Sect. \ref{Dust} for details.

% {\bf ***** IMPORTANTE, creo esto te lo dije en su momento: un objeto con esas Te y ne tiene abundancias de oxigeno de 8.30, para estas galaxias, todas de alta metalicidad, se espera que Te es mas baja, incluso por debajo de 5000 K, y ahi el cociente teorico Ha/Hb es 3.04, siguiendo Storey \& Hummer, 1995, MNRAS 272, 41, ?como afecta eso a las medidas que tienes? *********}. 

To work in the same system as the MPA-JHU Bayesian estimates used for the SDSS data, we  computed SFRs using Eq. \ref{SFRsHop} and \ref{LumHop}  for the SDSS data, and calibrated them to the \citet{Brinchmann04} SFRs system, obtaining:

\begin{equation}
{\rm log}({\rm SFR})_{\rm B04}=-0.0648 + 1.3759 \times {\rm log(SFR)}_{\rm Hopkins}
\end{equation} See Appendix 1 for more details. It is important to note that the estimated SFRs have been rescaled to a \citet{Kroupa01} IMF.

Finally, stellar masses were measured by  \citet{Taylor11},  who estimate the stellar mass--to--light ratio ($M_{\ast}$/L) from optical photometry  using  stellar population synthesis models. They demonstrate that the relation between ($g-i$) and $M_{\ast}$/L offers a simple indicator of the stellar masses. The stellar masses assume a \citet{Chabrier03} IMF.  It is worth noting that the \citet{Chabrier03} and \citet{Kroupa01} IMFs have very similar shapes, and that the conversion factor between them when estimating stellar masses is negligible \citep[e.g.][]{Haas10}.

% Is important to use the same IMFs for the SFR and stellar mass measurements to avoid inconsistencies in the analysis.

\subsubsection[]{Dust extinction correction} \label{Dust}

The extinction correction for our sample was derived using the \citet{Cardelli89} extinction law,  based on observed Balmer decrements assuming Case B recombination. The relation between the observed and corrected fluxes is given by $F(\lambda)_{\rm corr}=F(\lambda)_{\rm obs}10^{0.4*A_{\lambda}}$, where $F({\lambda)_{\rm corr}}$ and  $F({\lambda)_{\rm obs}}$ correspond to the corrected and observed fluxes, respectively.

Since GAMA is a fainter survey in magnitude than the SDSS, we expect, on average, larger extinction corrections. This comes about for two reasons. First, GAMA is sensitive to fainter systems than SDSS at similar redshift and luminosity, that are fainter due to being more heavily obscured. Second, GAMA is sensitive to high-SFR systems at higher redshift than SDSS, which have higher obscuration associated with their larger SFRs \citep[e.g.,][]{Hopkins03}. As we are deriving metallicities through relatively close emission line ratios ( [{O\,\textsc{iii}}]$\lambda$5007 / H$\beta$ and [{N\,\textsc{ii}}]$\lambda$6583 / H$\alpha$ ),  the correction for extinction essentially cancels out. An overestimation of the extinction for the H$\alpha$ line however, would result in an overestimation of the SFR,  which may mislead our conclusions about the evolution of the SFR.

\begin{figure}
\includegraphics[scale=0.6]{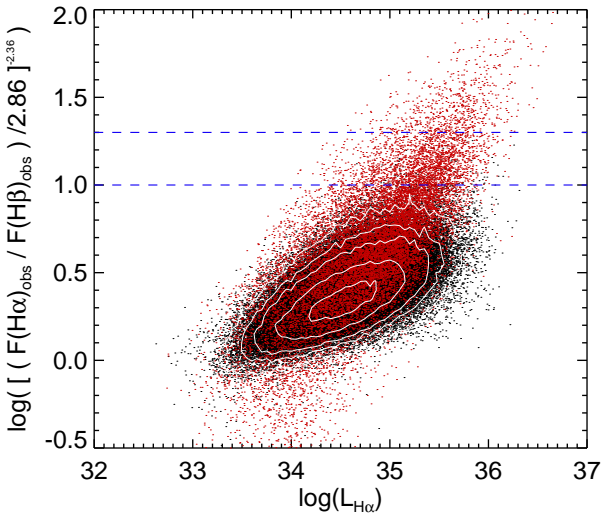}
\caption{H$\alpha$ Luminosity vs. reddening correction  factor  for SDSS (black) and GAMA (red) galaxies. Contours correspond to the SDSS sample. Blue dashed lines correspond to a reddening correction factor of 10 (Balmer decrement of 7.58) and 19.2 (Balmer Decrement of 10).}
\label{LumVsExt}
\end{figure}

To avoid any spuriously high Balmer decrements introducing overestimated SFRs in our analysis, we  imposed an upper limit of 10 to the obscuration correction, which results in a Balmer decrement of $\sim$7.58.  We  also compare our results taking an upper limit of $\sim$19.2 to the obscuration correction, which corresponds to a  Balmer Decrement of 10. Those limits are shown in Fig. \ref{LumVsExt}. The differences in SFR evolution are small, and will be studied in \S\,\ref{EvolutionS}.

\section[]{Evolution of the SFR, SSFR, Z, and \Ms\ relationships for GAMA and SDSS galaxies}\label{EvolutionS}

\begin{table*}
\begin{center}
\begin{tabular}{cc|c|cc||ccc||}
\hline
\hline
\multicolumn{8}{c} {Evolution (dex) }\\\cline{1-8}
\hline
Sample& redshift range  & \multicolumn{2}{|c|} {12+log(O/H)} & \multicolumn{2}{c} {log(SFR)}& \multicolumn{2}{c} {log(SSFR)} \\\hline

1	&	0.04 : 0.055	&	-	&		&	-	&		&-  &  \\\
2	&	0.055 : 0.07	&	-	&	V1: - &	 -	&	V1: - 	& - &	V1: - \\
3	&	 0.070 : 0.085	&	-	&		&	-	&		&-\\\
4	&	 0.085 : 0.1	&	-	&		&	-	&		&-\\\hline
5	&	 0.1 : 0.115	&	0.011	&	&	0.168	&	&	0.180	&	\\\
6	&	 0.115 : 0.130	&	0.016	&V2: 0.016 	&	0.231	& V2:  0.225	&	0.240	& V2 :  0.234	\\\
7	&	 0.130 : 0.145	&	0.022	&	&	0.252	&	&	0.266	&	\\\
8	&	 0.145 : 0.160	&	0.021	&	&	0.310	&	&	0.314	&	\\\hline
9	&	 0.160 : 0.175	&	0.035	&	&	0.392	&	&	0.398	&	\\\
10	&	 0.175 : 0.190	&	0.034	&V3: 0.049 	&	0.472	&V3:  0.459	&	0.552	& V3 : 0.542	\\\
11	&	 0.190 : 0.205	&	0.058	&	&	0.471	&	&	0.567	&	\\\
12	&	 0.205 : 0.235	&	0.066	&	&	0.443	&	&	0.571	&	\\\hline
13	&	 0.235 : 0.265	&	0.077	&	&	0.459	&	&	0.592	&	\\\
14	&	 0.265 : 0.295	&	 0.093	&V4: 0.088	&	0.456	&V4:  0.444	&	0.591	&  V4 : 0.581\\\
15	&	 0.295 : 0.330	&	0.083	&	&	0.419	&	&	0.564	&	\\\
16	&	 0.330 : 0.365	&	0.15	&	&	0.354	&	&	0.516	&	\\\hline
\noalign{\smallskip}
\end{tabular}
\normalsize
\rm
\end{center}
\caption{Evolution found in 12+log(O/H), SFR, and SSFR for the 16 volume limited samples,  and 4 concatenated samples.}
\label{Evolution}
\end{table*}

In this section we  use a statistically robust sample combining both SDSS and GAMA galaxies in volume limited samples up to z$\sim$0.36 to explore possible evolution of the \Ms$-Z$,  \Ms--{\it SFR},  \Ms--{\it SSFR}, and $Z$--{\it SFR} relationships.  
Figs. \ref{MZ} to \ref{MetSFR} plot these relationships for the 16 volume-limited samples described in $\S\,\ref{SampleSelection}$.  In these figures, data  from the SDSS are shown in black, while those from GAMA are shown in red. Both samples show good agreement in the internal consistency between the surveys, and in the overall trends. The large volume covered by the SDSS gives us a robust local fit for all our local relationships. For $z>$0.1 the SDSS survey becomes more incomplete for lower mass galaxies, however at this point, the GAMA sample fills in the lower mass populations. The evolution seen at high redshift  is evident primarily in the GAMA survey.

For GAMA galaxies, samples 8 and 9 (marked with a star from Figs. \ref{MZ} to \ref{MetSFR}) will not be taken into account because they are strongly affected by sky lines. The metallicity and SFR for these SDSS samples however, are not affected by sky line emission contamination because all emission lines  were used to derive these values. Therefore, we only consider data from the SDSS survey in samples 8 and 9.

To improve our statistical reliability within the subsamples, and to measure  the metallicity and SFR evolution more accurately, we  concatenated samples  with similar offsets in metallicity and SFR. Samples 1 to 4 form the sample V1; samples 5 to 8 (5 to 7 for GAMA) form sample V2; samples 9 to 12 (10 to 12 for GAMA) form sample V3; and samples 13 to 16 form sample V4.  The new samples V1 to V4 are shown  in the right panels  of Figs. \ref{MZ} to \ref{MetSFR}.  These samples will be used to analyze the dependencies between \Ms, SFR, and $Z$ (\S\,\ref{dependencies}) and to study the FP (\S\,\ref{FP}).

Since galaxies up to  $z\sim$0.1 do not show any sign of metallicity evolution in the \Ms$-Z$ relation and cover a wide range in stellar mass, we fit a second order polynomial to sample V1.  To get a robust fit, we are only considering galaxies between the mass limits shown by the vertical lines, and given in Fig. \ref{RedshiftVsMasa}.  The local fit to the \Ms$-Z$ relation is given by:
\begin{equation}\label{MZLocal}
12+{\rm log(O/H)}=-10.8297+3.6478x-0.16706x^2,
\end{equation} where  $x=\log(M_{\star}/M_{\odot})$, and $\sigma$=0.150.

This local \Ms$-Z$ relation is plotted in all the 16 samples shown in Fig. \ref{MZ}. To measure metallicity evolution  we consider as a single sample all galaxies from both the SDSS and GAMA surveys, as  shown in each panel of Fig. \ref{MZ}.  Then we fitted the zero point of Eq. \ref{MZLocal} for each sample. The resulting offset is shown  by a green line in each panel of Fig. \ref{MZ}, and the difference between the local and fitted zero point is given in Table \ref{Evolution}. In the case of the concatenated samples V1 to V4, the offset was also estimated taking into account both the GAMA and SDSS galaxies as a single sample.

From Fig. \ref{MZ} it is evident that the metallicity decreases with increasing redshift. The highest difference
between local and redshifted galaxies is found in sample 16  ($0.330<z<0.365$), which has an oxygen abundance which is $\sim$0.139~dex lower than that observed in local galaxies. The median error in metallicity of the concatenated samples are 0.048, 0.047, 0.047, 0.052 dex for samples V1, V2, V3, and V4, respectively. The median error is not only smaller than the possible evolution seen, but being a random error it is unlikely that the whole population would suffer a similar decrement randomly.

% with a maximum evolution of $\sim$0.139 dex in 12+log(O/H) for sample 16
%, see Fig. \ref{MZ} and Table  \ref{Evolution}.  %%% NO NECESARIO
%
A metallicity evolution at redshifts $z\leq 0.4$ was also found by \citet{Lara09a, Lara09b} and \citet{Pilyugin11}. In both studies the oxygen abundance shows a decline of up to $\sim$0.1~dex  between local and  $z\sim 0.4$  galaxies. This metallicity evolution  agrees with the predictions given by the models {of \citet{Buat08} for galaxies with a  velocity dispersion of $\sim$360~km\,s$^{-1}$ and log (\Ms/$M_{\odot}$) $\sim$ 11.25.

\begin{figure*}
\includegraphics[scale=0.62]{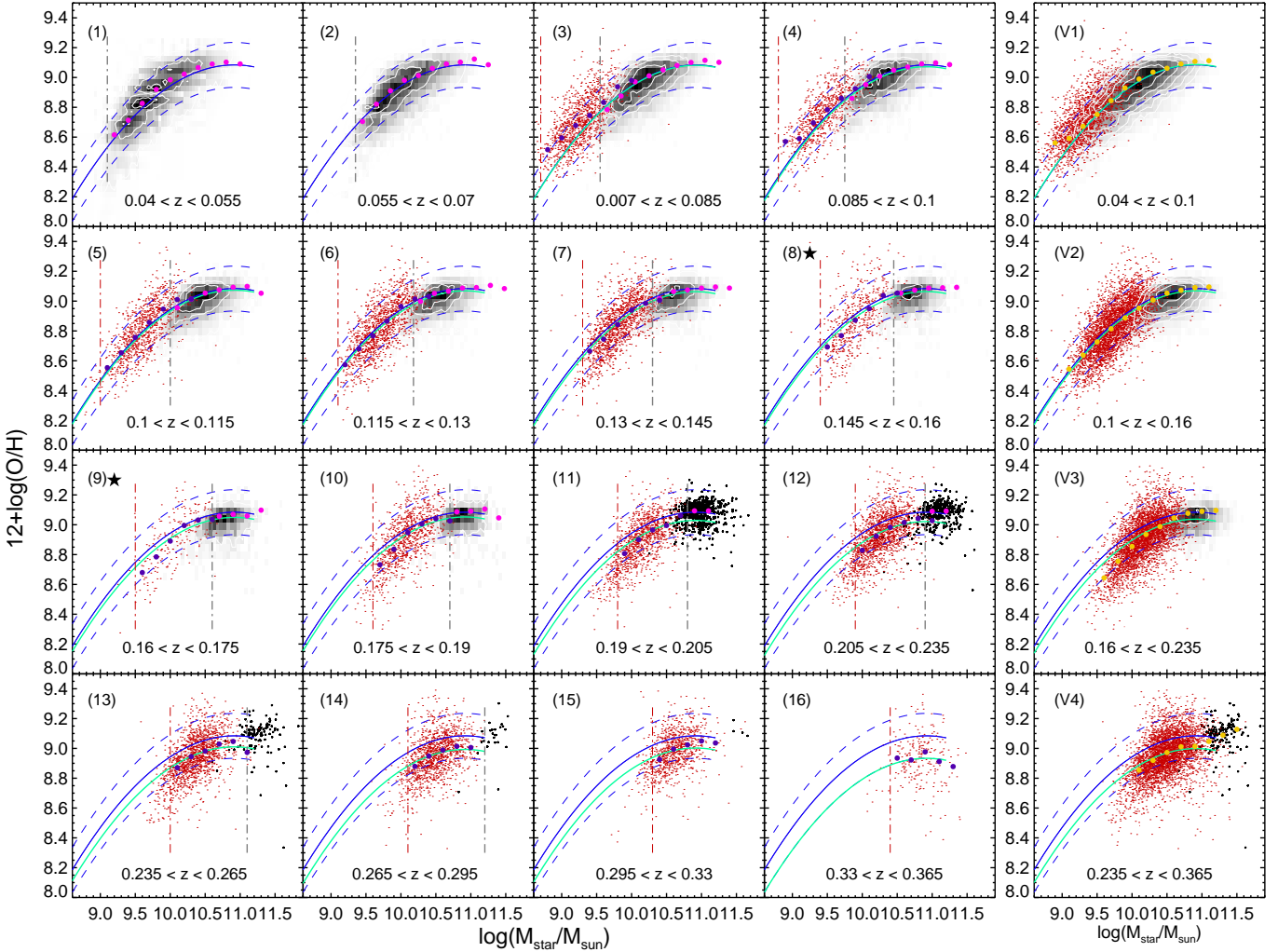}
\caption{$M_{\star}-Z$ relation for our volume-limited  galaxy samples using data from the GAMA (red dots) and SDSS (black dots) surveys. The  redshift range used is shown at the bottom of each panel, and the number in the top left corner of each panel identifies the volume limited sample used in each case. The star in samples 8 and 9 indicates the GAMA samples for which H$\alpha$ and [NII]~$\lambda$6584  are strongly affected by sky lines. The blue solid line, shown in all panels, corresponds to a 2nd order polynomial fit to the local SDSS and GAMA samples up to $z \sim$0.1  (Eq. \ref{MZLocal}). The blue dashed lines indicate the 1-$\sigma$ dispersion for this fit. The green solid line shows the result of fitting the zero-point in Eq. \ref{MZLocal} to each sample. Vertical dot-dashed lines correspond to the mass limits of SDSS (grey) and GAMA (red). Pink and purple circles correspond to the median metallicity in bins of stellar mass for SDSS, and GAMA, respectively.  The right panels show the concatenated samples V1 to V4, with  their respective redshift range at the bottom of each panel. The yellow circles indicate the median metallicity in bins of stellar mass taking GAMA and SDSS galaxies as a single sample.}
\label{MZ}
\end{figure*}

% AQUI COMIENZO A HABLAR DE LA MASA-SFR

In a similar way we generated the   \Ms--{\it SFR} relation for all our samples.  Fig. \ref{MSFR} shows this relation, which can be represented by a linear fit. The  \Ms--{\it SFR} relation for local galaxies up to $z \sim$ 0.1 (sample V1) is given by:

\begin{equation}\label{MSFRLocal}
{\rm log(SFR)}=-5.3126+0.5547x,
\end{equation} where  $x=\log(M_{\star}/M_{\odot}$), and $\sigma$=0.349

As we did in the case of the $M_{\star}-Z$ relation, we fitted the zero point of Eq. \ref{MSFRLocal} for each volume limited sample considering all galaxy data from GAMA and SDSS as a single sample. The difference between the local and the fitted zero point is  listed in Table \ref{Evolution}.  As can be seen, the SFR evolves to higher values as redshift increases, with a median evolution up to $z$=0.365 of $\sim$0.4 dex for sample V4.

As discussed in \S\,\ref{Dust}, we  consider  a  Balmer decrement of $\sim$7.58  as an upper limit to the dust correction to avoid overestimating the SFR in GAMA galaxies. There will be some systems that do have such high obscurations, however, that we are overlooking as a consequence. As a result, the evolution found  in log(SFR)  should be taken as a lower limit, in the event that some number of intrinsically high obscuration systems are being erroneously excluded.  Considering a more relaxed limit of 10 to the Balmer decrement, the evolution found for  log(SFR) in samples 13, 14, 15, and 16 is 0.05, 0.06, 0.05 and 0.07~dex greater, respectively, than the values listed in Table \ref{Evolution}.

\begin{figure*}
\includegraphics[scale=0.62]{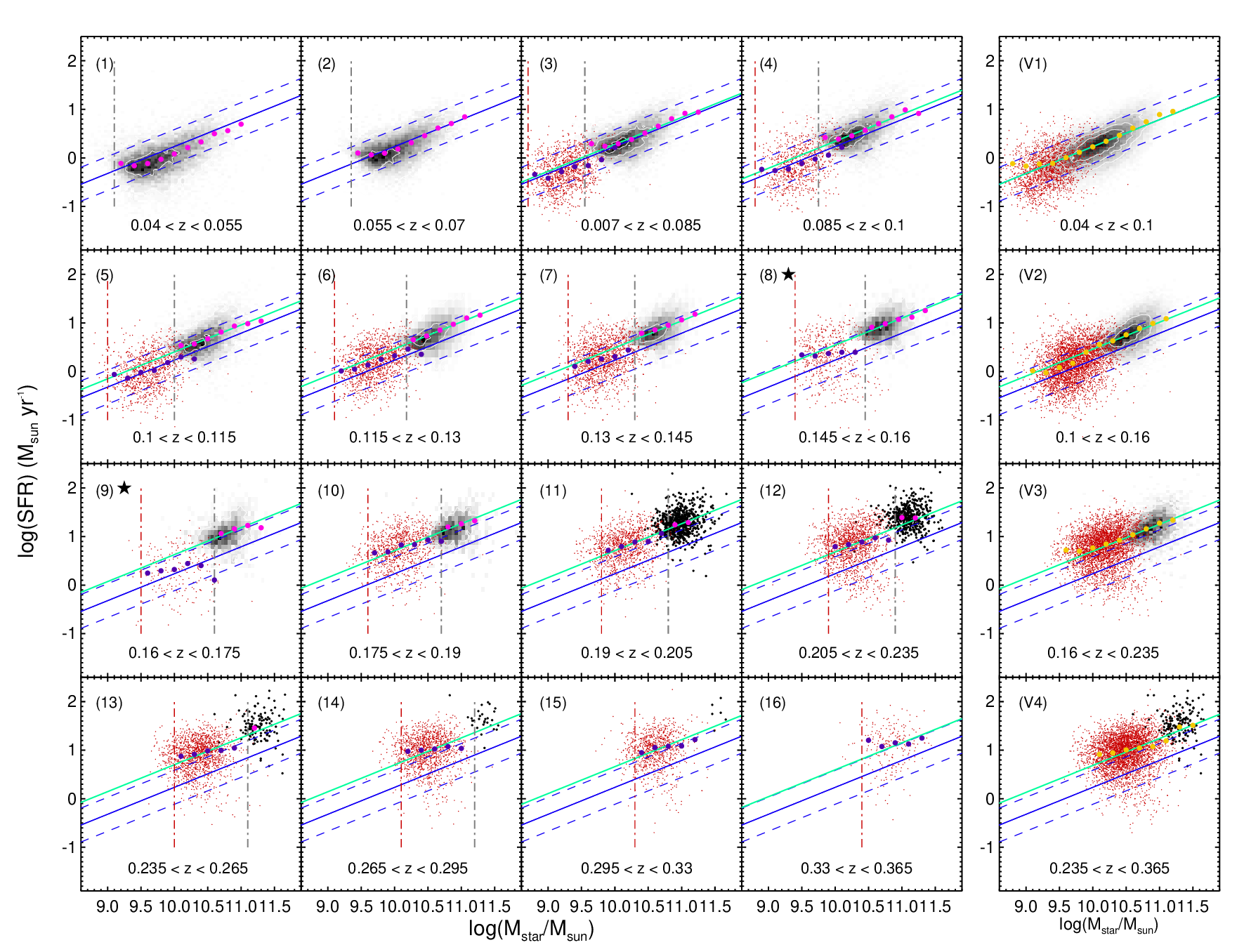}
\caption{As in Fig. \ref{MZ}, but now showing the $M_{\star}-{\it SFR}$  relation. Symbols, annotations and lines are as in Fig \ref{MZ}. The blue solid line, shown in all panels, corresponds to a linear fit to the local  samples up to $z \sim$0.1  (Eq. \ref{MSFRLocal}). The blue dashed lines indicate the 1-$\sigma$ dispersion for this fit. The green solid line shows the fit to the zero point of the local relation. Pink and purple circles correspond to the median SFR in bins of stellar mass for SDSS, and GAMA, respectively.  The yellow circles indicate the median SFR in bins of stellar mass taking GAMA and SDSS galaxies as a single sample.}
\label{MSFR}
\end{figure*}

% 
% SFR evoluton.
% The difference between taking galaixes with a upper Dust correction of 10 and 19 is 
% 
% Metalicidad evolution, V4: 0.071
% SFR evolution for sample V3  : 0.48359054
% SFR evolution for sample V4  : 0.47257099
% 
% SFR evolution for sample 13:  0.49047629
% SFR evolution for sample 14:  0.48778700
% SFR evolution for sample 15:  0.42685264
% SFR evolution for sample 16:  0.44759512

We also analyze the   \Ms--{\it SSFR} relation because it offers an excellent way of studying the effect of galaxy  downsizing. Figure \ref{MSSFR} shows the analysis of the   \Ms--{\it SSFR} relation.  We find results consistent with earlier work of the  \Ms--{\it SSFR} \citep[e.g. Bauer et al. 2012,][]{Noeske07a}. A linear fit gives a good representation  of the   \Ms--{\it SSFR} relation for these samples of star forming galaxies. The  \Ms--{\it SSFR} relation for the local  (sample V1, $z \lesssim$ 0.1)  is given by:

\begin{equation}\label{MSSFRLocal}
{\rm log(SSFR)}=-5.479-0.434x
\end{equation} where  $x=\log(M_{\star}/M_{\odot})$, and $\sigma$=0.304.

We again fit the zero point of Eq. \ref{MSSFRLocal} for each volume limited sample. The difference between the local and the fitted zero point is listed in Table \ref{Evolution}.

The high SSFR seen for low mass galaxies indicates they are increasing their stellar mass, relatively speaking, more quickly than those at high mass, \citep[e.g.][]{Noeske07a}. We measure a change in the zero point as large as the scatter in the relation over the $0\lesssim z \lesssim 0.36$ span explored here. We find a evolution of 0.56~dex for sample V4 (see Table  \ref{Evolution}).

\begin{figure*}
\includegraphics[scale=0.62]{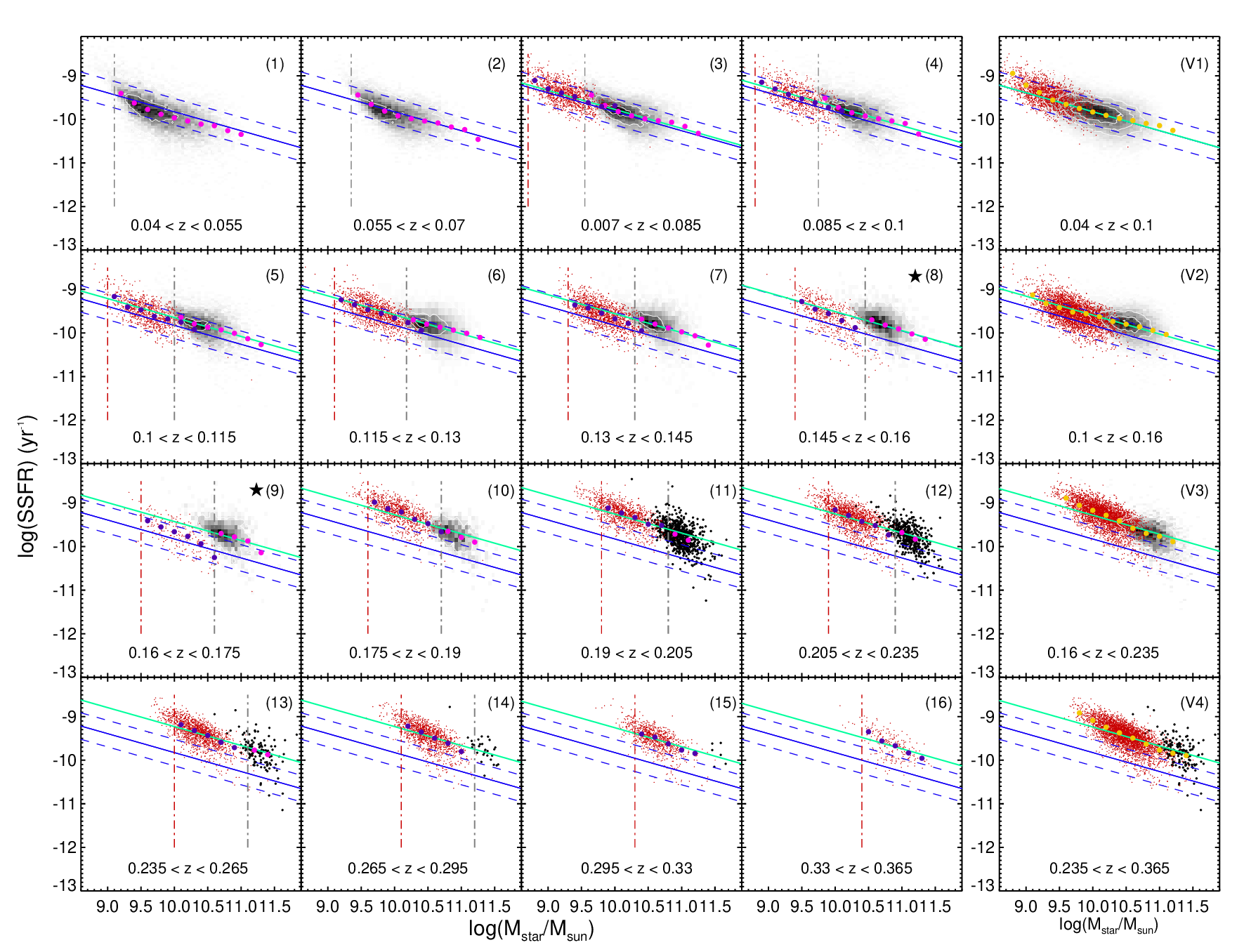}
\caption{As in Fig. \ref{MZ}, but now showing the $M_{\star}-{\it SSFR}$  relation. The blue solid line, shown in all panels, corresponds to a linear fit to the local  samples up to $z \sim$0.1  (Eq. \ref{MSSFRLocal}). The blue dashed lines indicate the 1-$\sigma$ dispersion for this fit. The green solid line shows the fit to the zero point of the local relation. The yellow circles indicate the median SSFR in bins of stellar mass taking GAMA and SDSS galaxies as a single sample.}
\label{MSSFR}
\end{figure*}

Finally,  Fig. \ref{MetSFR} studies  the $Z$--{\it SFR} relation. As already discussed, this relation has the broadest scatter of all  combinations of SFR, \Ms\ and $Z$.  A third-order polynomial fit to the V1 sample yields

\begin{equation}\label{MetSFRLocal}
\log({\rm SFR})=-1443.97+508.48x-59.692x^2+2.336x^3,
\end{equation} where  $x$=12+log(O/H), and $\sigma$=0.2.
Since this relation  has a very high dispersion and both variables,  $Z$ and SFR, evolve with redshift, we do not measure any evolution in this relationship. However we observe a general tendency for the SFR to increase with metallicity for all the volume limited samples.  We  discuss the  $Z$--{\it SFR} relation in detail in \S\,\ref{dependencies}.

\begin{figure*}
\includegraphics[scale=0.62]{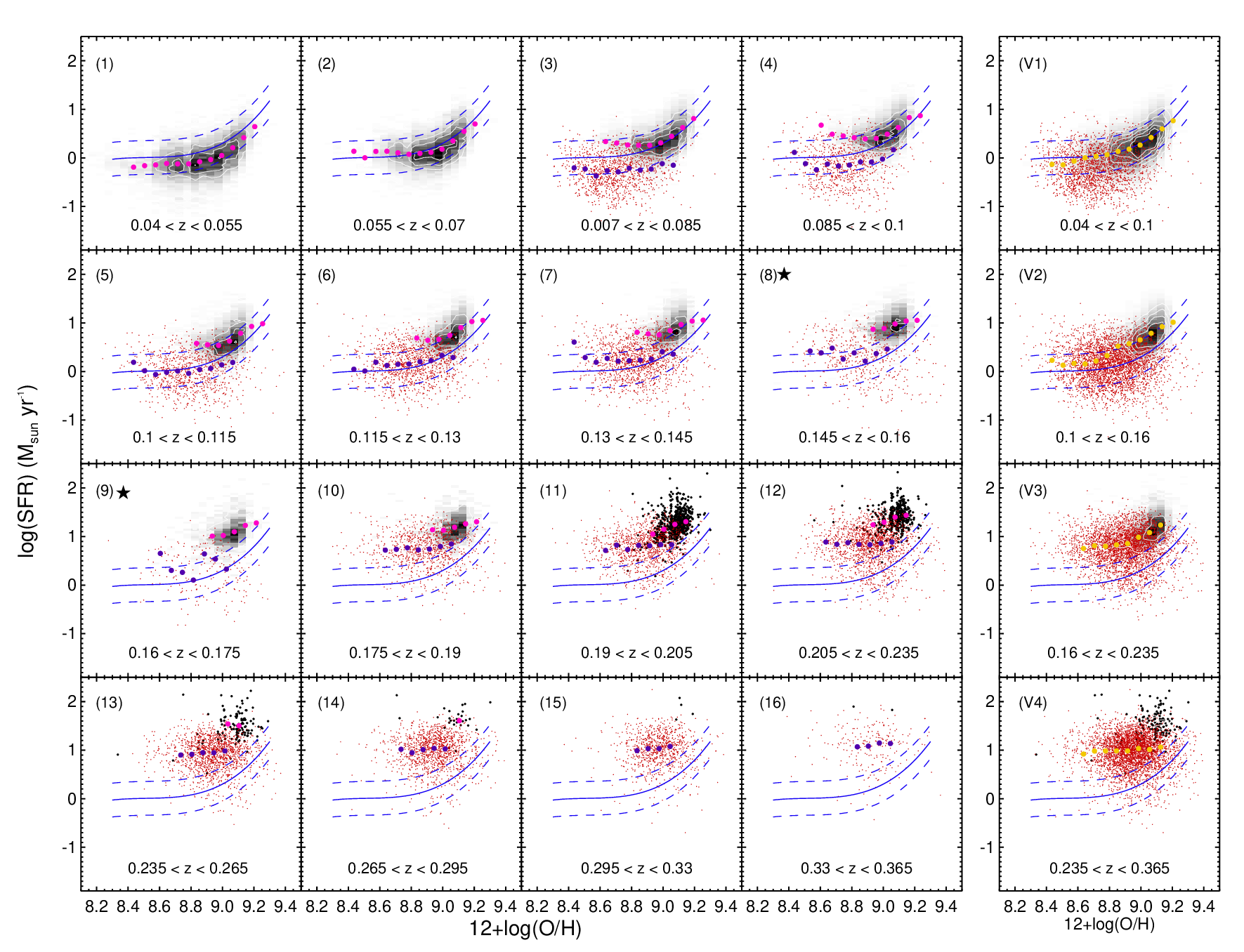}
\caption{As in Fig. \ref{MZ}, but now showing the $Z$--{\it SFR}  relation. The blue solid line, shown in all panels, corresponds to a 3rd order polynomial fit to the local  samples up to $z \sim$0.1  (Eq. \ref{MetSFRLocal}). The blue dashed lines indicate the 1-$\sigma$ dispersion for this fit. The yellow circles indicate the median SFR in bins of $Z$ taking GAMA and SDSS galaxies as a single sample.}
\label{MetSFR}
\end{figure*}

% ****************************

\section[]{The Fundamental Plane}\label{FP}

We generate the FP considering all the volume-limited samples for the GAMA and SDSS galaxies as a single sample. The \Ms$-Z$,  \Ms--{\it SFR}, and $Z$--{\it SFR} relationships are the  projections of this 3D distribution. While \Ms\ correlates with both SFR and metallicity (the well known \Ms$-Z$ and  \Ms--{\it SFR} relationships), the SFR does not  strongly correlate with metallicity (see Fig. \ref{MetSFR}), which means that this relation is close to the face-on view of the 3D distribution (see top left panel of Fig. \ref{Cubos3D}).

\citet[][]{Lara12} explore different methods to analyze and give the best representation to the \Ms--$Z$--{\it SFR} space. The methods analysed include principal component analysis (PCA), regression, and binning data. The result that best quantifies the distribution of galaxy measurements in this space is by fitting a plane to the stellar mass using regression. Although PCA does not give the best fit, a PCA analysis for our GAMA and SDSS sample indicates that the first two principal components account for  98$\%$ of the variance, confirming that a planar representation is appropriate.

Following \citet{Lara10a} and \citet{Lara12},  we fitted a plane to \Ms\ using regression:

{ \small \begin{equation}\label{FPEq}
{\rm log}({M_{\star}}/{M_{\odot}}) = \alpha \ [12+\rm log(O/H)] + \  \beta \ [log(SFR)] \ + \gamma,
\end{equation}} 
where $\alpha$=1.3764 $\pm$ 0.006, $\beta$=0.6073 $\pm$ 0.002, and  $\gamma$=-2.5499 $\pm$ 0.058.
The FP given by Eq. \ref{FPEq}, which provides a good approximation to our data, is shown in Fig. \ref{Cubos3D}. This relation recovers the \Ms\ of our entire sample  with $\sigma$=0.2~dex.  As  demonstrated by \citet{Lara10a}, this FP can also recover the \Ms\ of high redshift galaxies,  and  shows a lack of evolution up to $z \sim 3.5$. See \citet{Lara10a} for a detailed discussion.

\begin{figure*}
\includegraphics[scale=0.5]{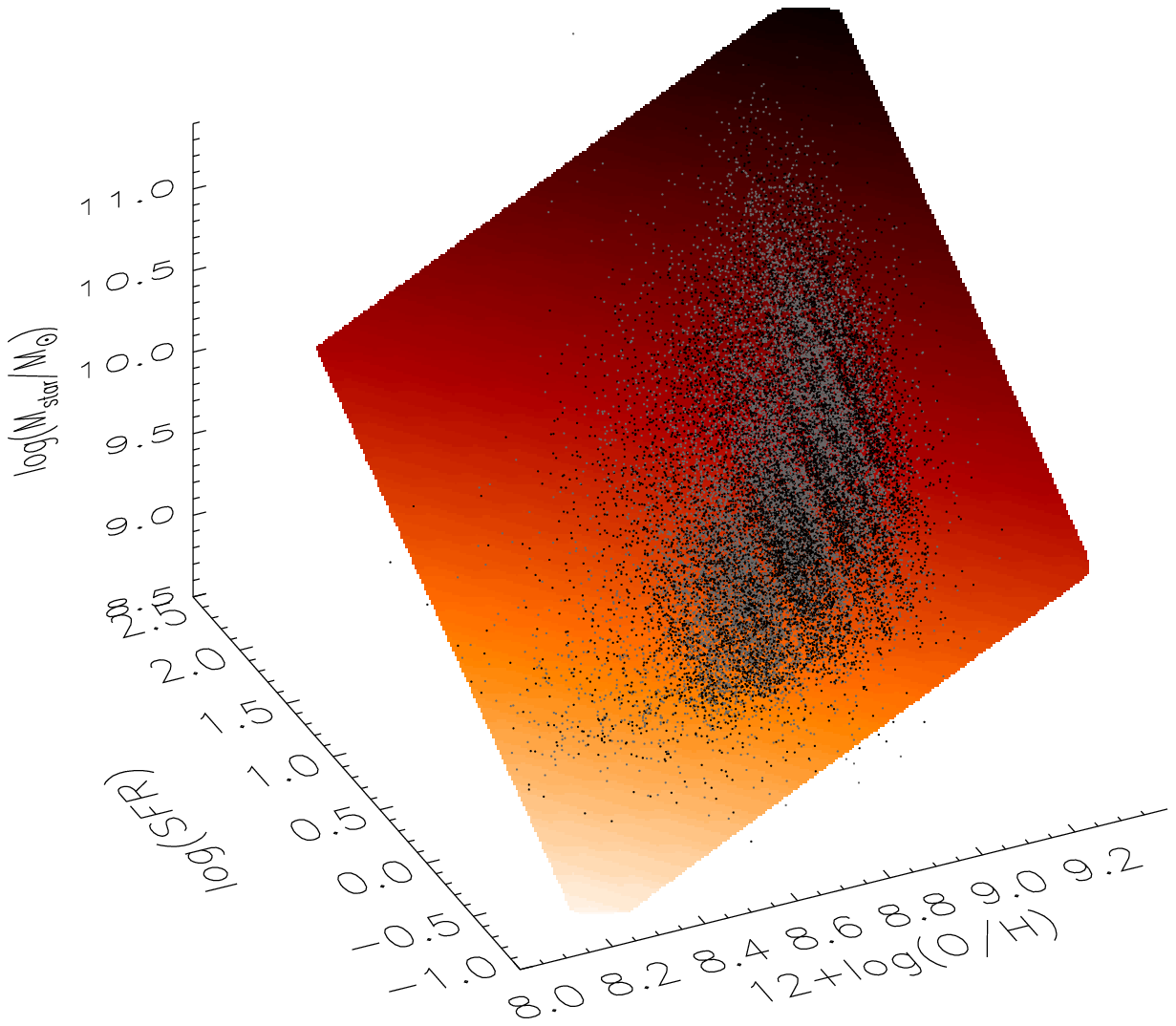}
\includegraphics[scale=0.5]{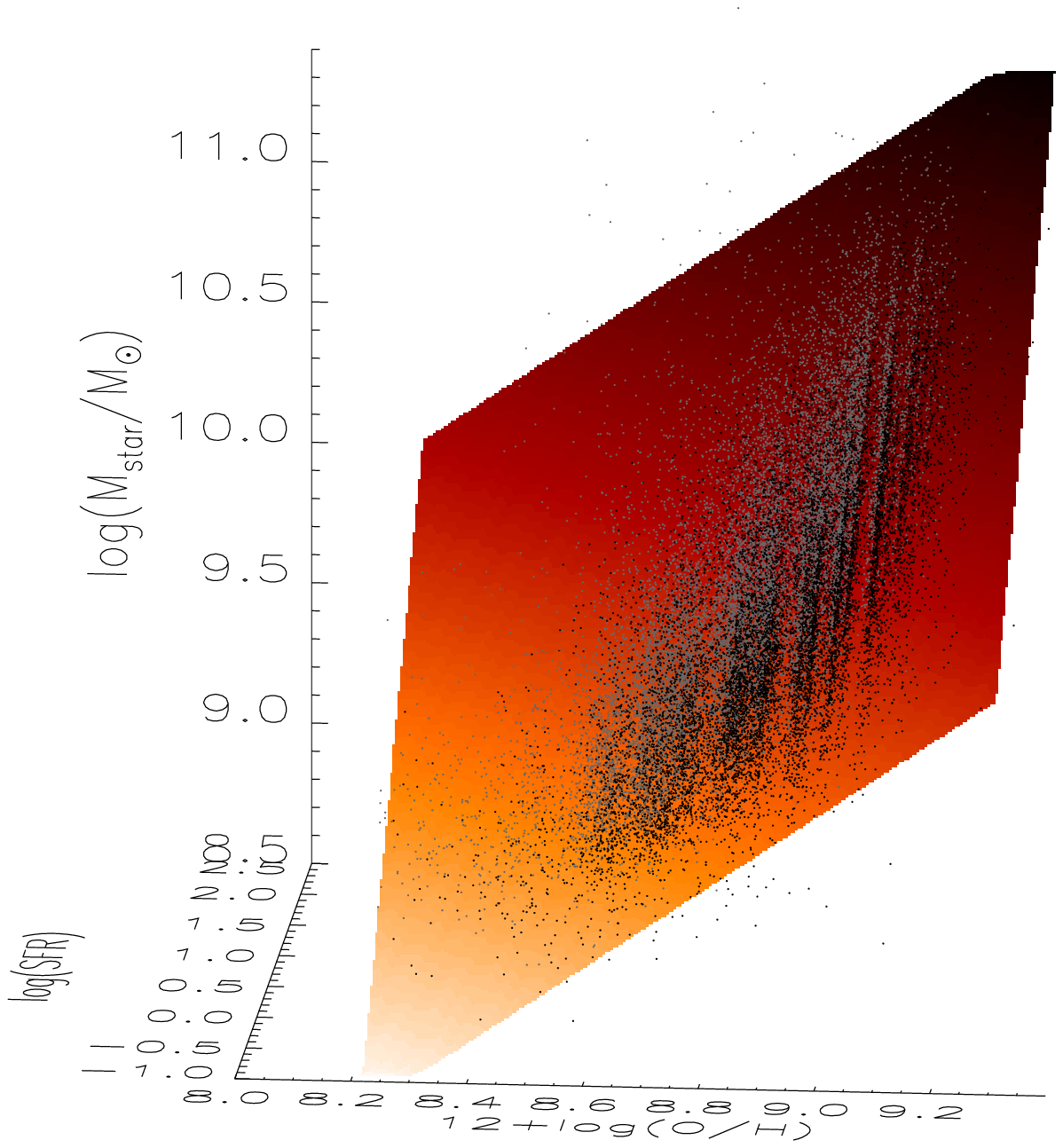}
\includegraphics[scale=0.5]{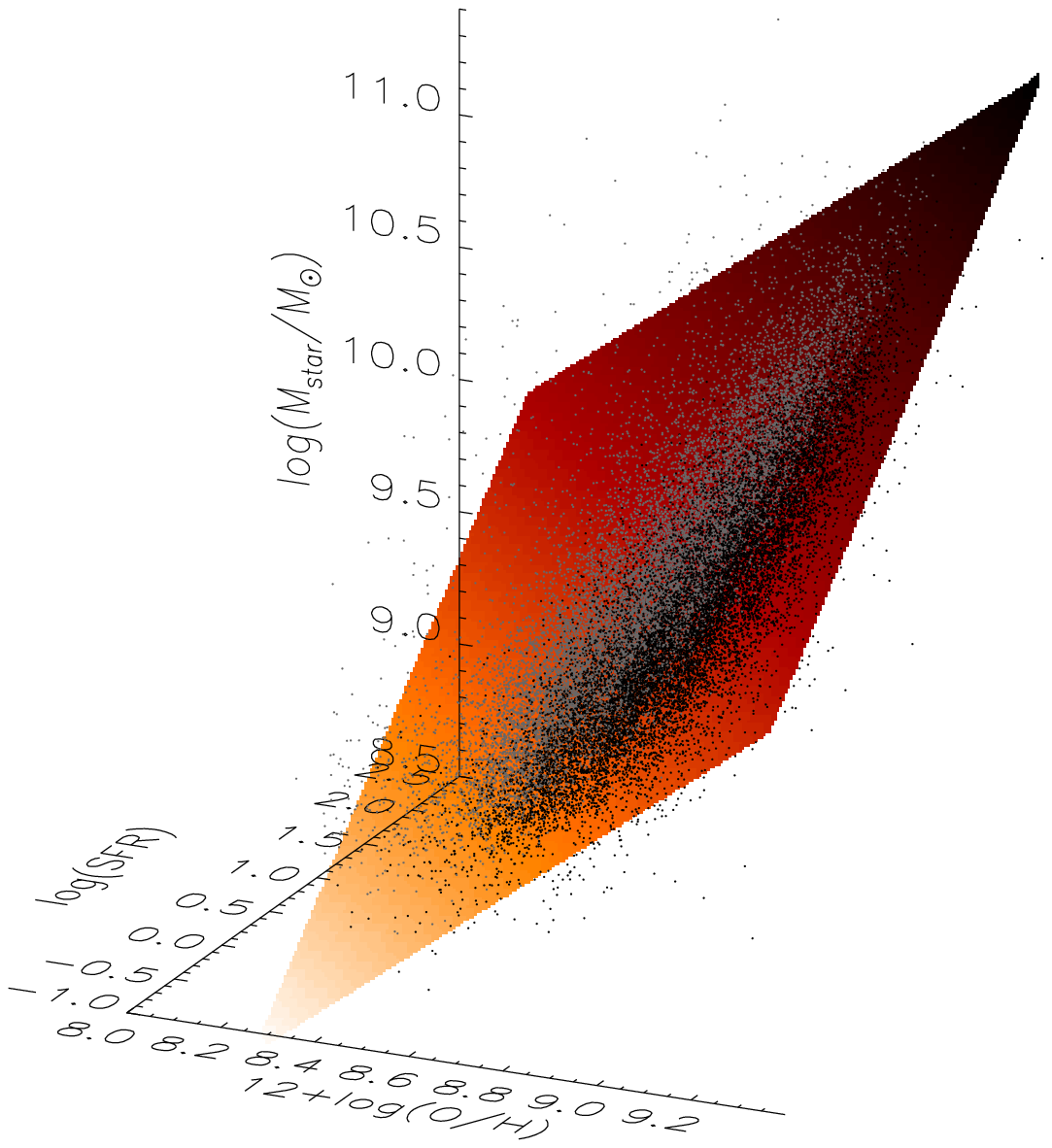}
\includegraphics[scale=0.5]{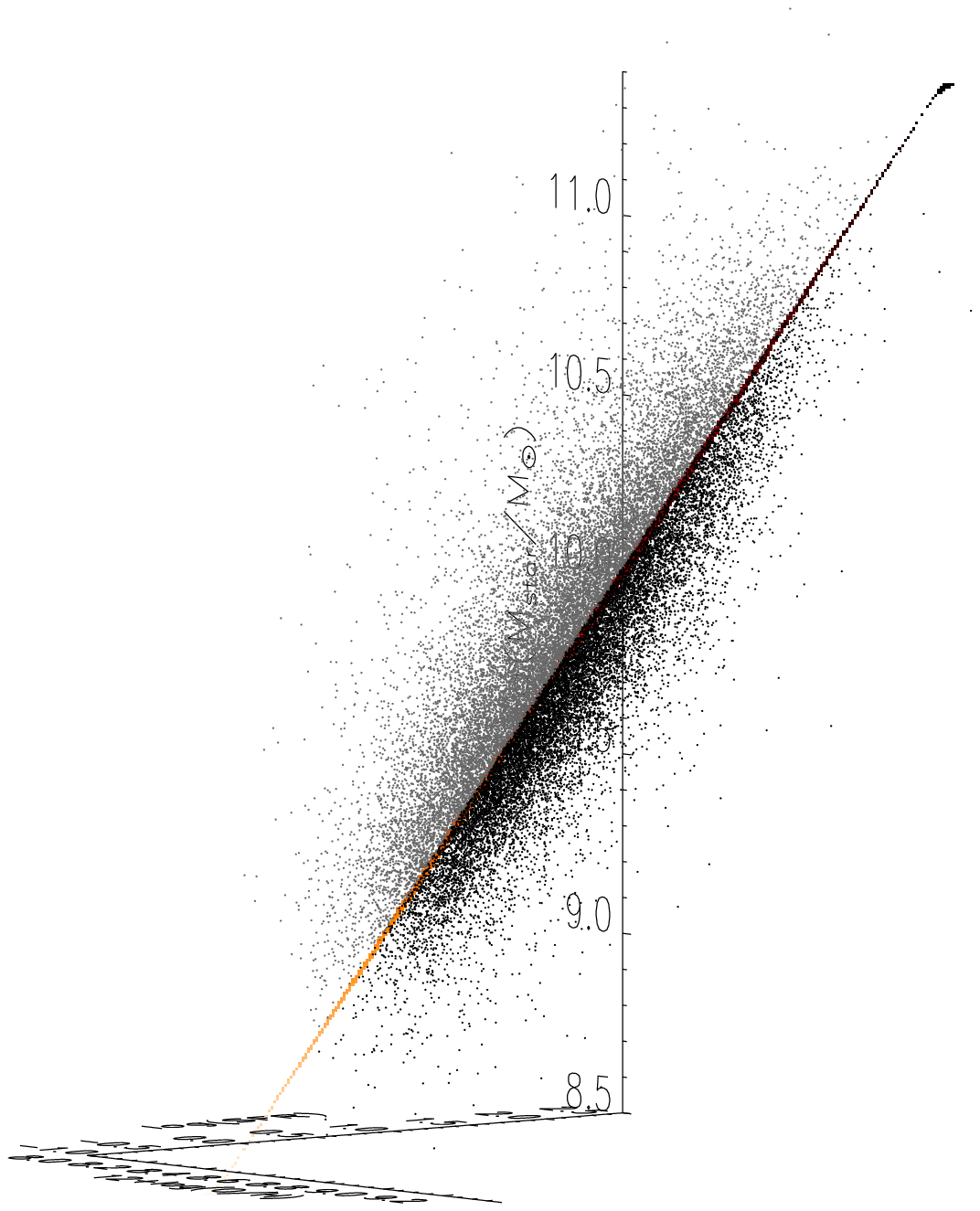}
\caption{Projections of the 3D distribution formed by  \Ms, log(SFR), and 12+log(O/H) for GAMA and SDSS galaxies. The orange plane shows the FP described in Eq. \ref{FPEq}.  The vertical axis shows  \Ms\ in all panels. The cube is rotated clockwise from the upper-left to the bottom-right panel. This last panel shows the edge-on projection of our derived FP. Grey and black dots show galaxies above and below the FP, respectively.}
\label{Cubos3D}
\end{figure*}

\begin{figure*}
\includegraphics[scale=1.0]{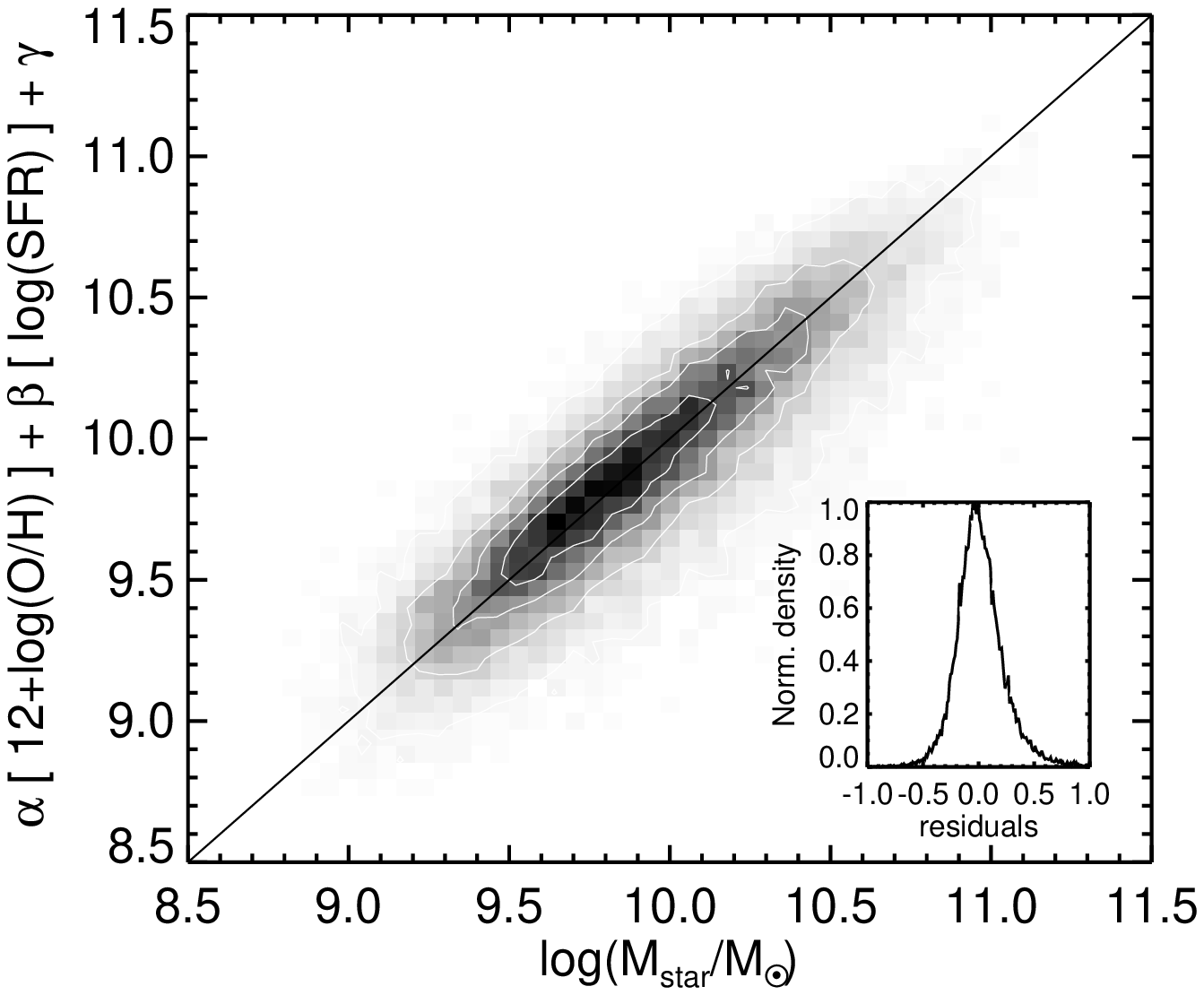}
\caption{Projection of the FP  for the GAMA and SDSS galaxies. The horizontal axis shows the observed \Ms, while the vertical axis shows the \Ms\ estimated through the FP described in Eq. \ref{FPEq}. The enclosed panel shows the histogram of the residuals with $\sigma$ = 0.2 dex.}
\label{FPP}
\end{figure*}

The FP is a consequence of the tight dependence of  \Ms\ on  SFR and $Z$. The current mass locked up in stars in a galaxy (\Ms) is a measure of the amount of gas currently being converted into stars (SFR), plus a measure of the star formation history, or past generations of stars, here represented by the metallicity ($Z$). So in broad terms, the value of \Ms\ can be though of as being dependent on a combination of the current SFR and the star formation history of a galaxy.

% HABLAR FISICAMENTE SOBRE LA FALTA DE EVOLUCION, LA MASA ES SFR+Z, 

To explain the lack of evolution in the FP  a description of how galaxies evolve is needed. 
At high redshifts, (proto-) galaxies  are characterized by stars of first generation that have not yet processed their gas. This implies low gas metallicities and very high SFRs and SSFRs. However, galaxies in the local universe  are characterized for stars of later generations that have been formed from pre-enriched gas and  reached higher gas metallicities. As these galaxies have a large amount of gas locked up into stars they show a lower SFR than those galaxies observed in the primitive universe. Hence,  there is an equilibrium between $Z$ and SFR at all redshifts. 
At high redshifts, the SFR is the fundamental parameter that drives the evolution of the galaxies. In the local universe however, $Z$ is the dominant parameter as the result of the gas processed into stars.  SFR and $Z$ are evolving in opposite directions:
while high-redshift galaxies show high SFRs and low $Z$, low-redshift galaxies host low SFRs and high metallicities. 
Hence, a linear combination of $Z$ and SFR will tend to cancel out  evolution  of the FP with redshift.  The evolution of the FP was studied by \citet{Lara10a}, who show that it does not evolve up to $z\sim$3.5.

It is noteworthy that although we are finding evolution in $Z$ and SFR as redshift increases, this is not in contradiction with the absence of evolution found in the FP. The FP suggests a new way of interpreting the stellar mass as a dependent property of both, $Z$ and SFR. The fact that the same FP can be use to recover the stellar mass of a galaxy at high redshift is a consequence of the evolution in opposite directions of the $Z$ and SFR, as mentioned above.

\section[]{SFR, SSFR, metallicity, and Stellar Mass dependencies}\label{dependencies}

To understand the properties of a pair of variables as a function of a third one, we perform a 2D analysis of the 4 variables we have been working on. For the volume limited samples V1 to V4, taking into account both SDSS and GAMA galaxies as a single sample, we study the SFR and SSFR dependence on the \Ms$-Z$ relation, the $Z$ dependence on the  \Ms--{\it SFR} and  \Ms--{\it SSFR} relations, and the \Ms\ dependence on the $Z$--{\it SFR} and $Z$--{\it SSFR} relations (see Figs. \ref{DependenciaBin2} to \ref{DependenciaBin1p1}).

% However, it should be emphasized that when working in 3 dimensions all the 3 variables must be taken into account in a 3D space, otherwise, a 2D projection of the variables onto a 3D space, could lead us to improper structures.

For these relationships, we can use the median values of one variable within bins of the other two to look at overall trends. When doing this, however, care must be taken to correctly interpret the results. For example, when studying the SFR dependence of the \Ms$-Z$ relation, we obtain  results that at first glance appear different if we estimate the median \Ms\ in  bins of metallicity (vertical bins, see Fig. \ref{DependenciaBin1}a), or the median Z in  bins of stellar mass (horizontal bins, see Fig. \ref{DependenciaBin2}a). These results are not contradictory, but provide different information. They should be carefully interpreted as saying, in the former case, how the median {\em stellar mass\/} varies for a given SFR and metallicity,  and in the latter how the median {\em metallicity\/} varies for a given SFR and stellar mass. These are clearly not the same thing, although in a diagram such as this it is easy to confuse the interpretation. To disentangle the different information that each binning direction  provides, we analyze  both. 

Figures \ref{DependenciaBin2} and \ref{DependenciaBin2p1} show  the combinations of the different dependencies of \Ms, $Z$, SFR, and SSFR. In these figures we take bins of the horizontal variable and estimate the median value of the vertical variable in bins of the third. Figures \ref{DependenciaBin1} and \ref{DependenciaBin1p1} show exactly the same distributions, but now binning in the other  direction.  Here we performe the binning in the vertical variable and estimate the median value of the horizontal variable in bins of the third. In each panel, the vertical and horizontal color lines show the 1-$\sigma$ dispersion of four representative bins of the variable shown in the color bar.

\subsection[]{Horizontal bins: estimating Z, SFR, and SSFR}\label{SFRonMZhorizontal}

Figure \ref{DependenciaBin2}a  shows the \Ms$-Z$ relation taking $Z$ as the principal variable to determine, the median $Z$ in bins of \Ms\ and SFR. We see a reversal at $\log(M/M_{\odot})\sim$10.2 in the SFR dependence of the metallicity of galaxies. For $\log(M/M_{\odot})\gtrsim$10.2, galaxies with high SFRs  have higher metallicities than galaxies with lower SFRs. On the contrary, for $\log(M/M_{\odot})\lesssim$10.2, galaxies with high SFRs show lower metallicities than galaxies with lower SFRs. As redshift increases, we still observe the same tendency for samples V2 to V3, and although the sample V4 does not span a broad range of \Ms,  our data suggest a similar trend. Sample V4 shows the metallicity evolution discussed in the previous section. Galaxies with a high SFR (pink circles in Fig \ref{DependenciaBin2}a) have metallicities which are $\sim$0.07~dex  lower than that observed in sample V1.  Considering the SSFR dependency (Fig. \ref{DependenciaBin2}b) a similar behaviour is found for $\log(M/M_{\odot})\lesssim$10.2, in the sense that galaxies with higher SSFR have lower metallicities than galaxies with a lower SSFR, and vice-versa for higher mass systems.

Figure \ref{DependenciaBin2}c and \ref{DependenciaBin2}d show the \Ms--{\it SFR} and \Ms--{\it SSFR} relationships taking the median SFR (and SSFR) in bins of \Ms\ and $Z$. In this case, we  observe  again the same reversal shown in for Fig. \ref{DependenciaBin2}a,b. 
At the high-mass end of the \Ms--{\it SFR} (--{\it SSFR}) relation, galaxies with  high metallicity show  higher SFR (and SSFR) than galaxies with  lower metallicity. 
On the other hand, at the low-mass end of the \Ms--{\it SFR} ({\it SSFR}) relation, galaxies with a high metallicity show lower SFR (and SSFR) than low metallicity galaxies.

To explain this behaviour, imagine two galaxies at the same redshift with similar stellar masses ($\log(M/M_{\odot})\sim11~M_{\odot}$), but one with a higher amount of neutral  gas than the other. Since both galaxies are massive, downsizing indicates that they are going to process their gas quickly.  Then, the galaxy with a larger amount of gas will reach higher metallicities and will have a higher  SFR and SSFR than the galaxy with less neutral gas.

On the other hand,  we now consider two low-mass  galaxies ($\log(M/M_{\odot})\sim9.5~M_{\odot}$) but one hosting  more neutral gas than the other. Downsizing indicates that, due to their low-stellar mass, both galaxies will process their gas slowly and on longer timescales than massive galaxies. According to Fig. \ref{DependenciaBin2}b, low-mass galaxies with a high SSFR show lower metallicities than galaxies with a lower SSFR.  The high metallicity of low mass galaxies  can be explained for a more bursty star formation in the past that exhausted its gas and increased its metallicity.

% Since higher SSFRs should enrich the gas quickly, the lower metallicities are likely to be due to galactic winds driven by the combination of relatively high SFR and low stellar mass.

From these considerations we infer that  both the amount of baryonic mass (stars plus gas) and downsizing are driving the  rate at which a galaxy is producing their metals. Massive galaxies with a large amount of gas will process their gas faster and reach higher metallicities than a similar galaxy with a smaller amount  gas. A  low-mass galaxy with a large amount of gas however, will tend to have a very low star formation efficiency, resulting in lower metallicities than a similar galaxy with lower gas, which have experimented a bursty star formation in the past. A detailed model of this picture is given in \citet{Lara13}.

% Z-SFR relation

% The Z-SFR and Z-SSFR, shown in Figs. \ref{DependenciaBin2}e and f  are special relations, and we will show that depending of the binning direction used, we will obtain completely different tendencies of the median data. For one hand, we have that the general tendency is of SFR increasing as Z increases. However, when the 

\subsection[]{Vertical bins: estimating  \Ms}\label{SFRonMZVertical}

Now we consider  stellar mass as the principal variable to be determined as a function of the others. Figure \ref{DependenciaBin1}a shows the  \Ms$-Z$ relation derived taking the median \Ms\ in bins of Z and SFR. The median values in this relation show that at a given metallicity, galaxies with higher SFR have higher median  \Ms.

It is possible to project the FP over the  \Ms$-Z$ face of the 3D-cube by giving values to the SFR and $Z$ and then estimate the \Ms\ through Eq.~(\ref{FPEq}).  Following this procedure we obtain the projections of the FP shown in Fig. \ref{DependenciaBin1}a, which  matches the median \Ms\ values found before.

Since the SSFR is a more physical quantity showing the actual amount of stars formed per unit mass, we show the SSFR vs. $Z$ in Fig. \ref{DependenciaBin1}b. In this case we see that low mass galaxies have low Z and high SSFR. Indeed, at any Z, higher SSFR galaxies are seen to have lower mass than those with lower SSFR.

% low-metallicity galaxies show a higher SSFR than galaxies with higher metallicities. This implies that low-mass galaxies are actively forming stars while processing their gas, and thus they have not yet reached high metallicities. 
% On the contrary, high-mass galaxies show the lowest SSFR and highest metallicities, 
% confirming that massive systems  are finishing the assembling of their mass and  processed the majority of their available gas. Therefore they reach high metallicities,  which is in agreement with the downsizing scenario. 

Figure \ref{DependenciaBin1}c shows the \Ms--{\it SFR} relation as a function of metallicity  taking the median \Ms\ in bins of SFR and $Z$. For a given SFR, higher metallicity systems have higher stellar mass,  i.e., we are obtaining the  \Ms$-Z$ relation. 
The solid lines in Fig. \ref{DependenciaBin1}c show the projection of the FP over this  \Ms--{\it SFR} relation and confirm that, at fixed metallicity, the median \Ms\ is higher for higher SFR.

Finally, Fig. \ref{DependenciaBin1}d studies the \Ms--{\it SSFR} relation as a function of metallicity. Again, at a given SSFR, higher metallicity galaxies have higher median stellar mass. Similarly, at a given metallicity, higher SSFR systems have lower median stellar mass.

Together, all these approaches to exploring the distribution of \Ms\, SFR and $Z$ provide a consistent picture of galaxy properties. In particular, the reversal seen in Fig. \ref{DependenciaBin2} in the dependence of $Z$ on SFR or SSFR as a function of mass is tantalisingly sugestive of a key role being played by the gas reservoir available for forming stars.

Any evolution  of the $Z$ or SFR will be along the projection of the plane. For example, sample V4 of Fig. \ref{DependenciaBin1}a shows an evolution toward lower values of $Z$. However, this evolution  actually comes from  the projections of the FP on the \Ms$-Z$ relation. On the other hand, sample V4 of Fig. \ref{DependenciaBin1}c shows a SFR evolution toward higher values of SFR. This evolution is again happening along the projections of the FP.

\begin{figure*}
\includegraphics[scale=0.87]{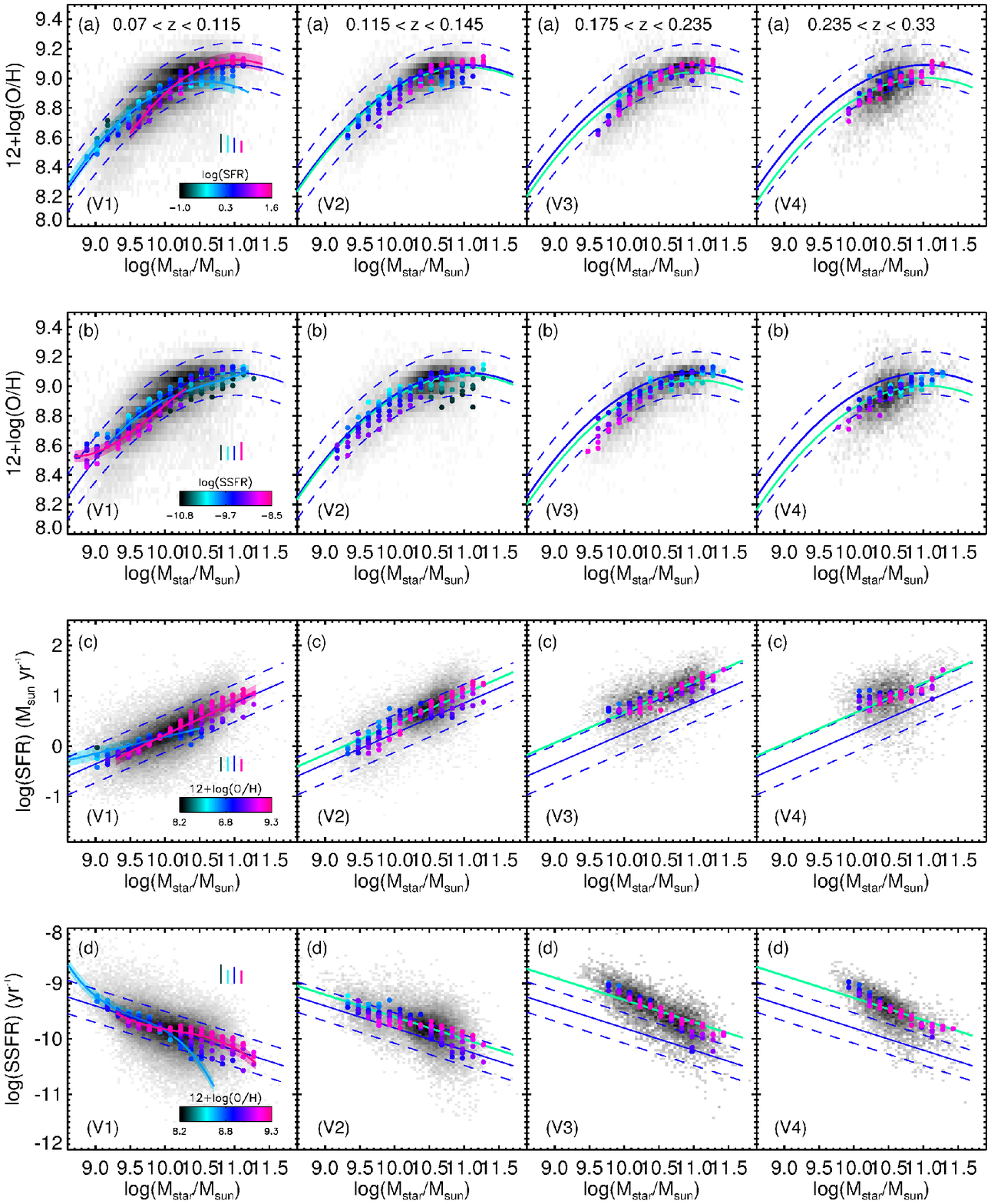}
\caption{From top to bottom, the SFR dependence of the \Ms$-Z$ relation, the SSFR dependence of the \Ms$-Z$ relation, the $Z$ dependence of the  \Ms--{\it SFR} relation, and  the $Z$ dependence of the  \Ms--{\it SSFR} relation. In all panels, we show the median value of the vertical variable in  \Ms\ bins (horizontal bins) for every bin of the third variable. The pink and blue ribbons in the left panels highligh the the median circles of the high and low mass end, respectively.  The vertical color lines show the 1-$\sigma$ dispersion of the vertical variable of four representative bins, as shown in the color-bar of each panel. The data density corresponds to the SDSS and GAMA sample together.}
\label{DependenciaBin2}
\end{figure*}

\begin{figure*}
\includegraphics[scale=0.87]{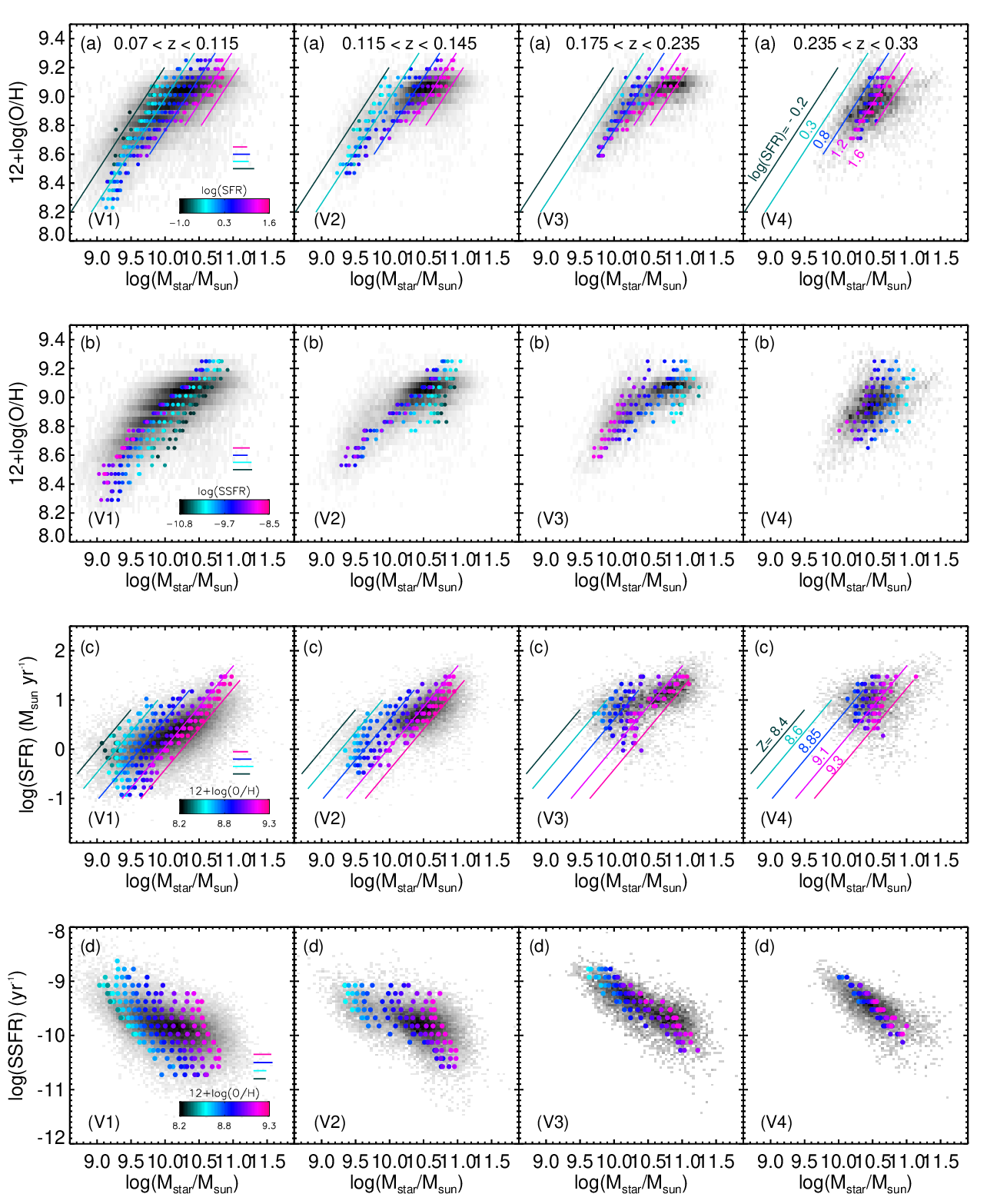}
\caption{From top to bottom, the SFR dependence of the \Ms$-Z$ relation, the SSFR dependence of the \Ms$-Z$ relation, the $Z$ dependence of the  \Ms--{\it SFR} relation, and  the $Z$ dependence of the  \Ms--{\it SSFR} relation. In all panels, we show the median \Ms\ in bins of the vertical variable (vertical bins) for every bin of the third variable.  The horizontal color lines show the 1-$\sigma$ dispersion of the horizontal variable of four representative bins, as shown in the color-bar of each panel.  The data density greyscale corresponds to the SDSS and GAMA sample together.}
\label{DependenciaBin1}
\end{figure*}

% The mean of the four 1-$\sigma$ dispersion of the metallicity median points shown in Fig 12a gives $\sim$0.1 dex. Whereas the mean of the four 1-$\sigma$ dispersion of the \Ms\ median 

% around a bin centered in log(SFR)$\sim$0.3, is 0.1 dex in metallicity (Fig 12a) and 0.2 dex in stellar mass (Fig 13a). 

 Finally, we examine the typical dispersions of vertical and horizontal bins in the  \Ms$-Z$ relation. In order to compare the degree of dispersion of two different variables, we use the coefficient of variation $C_v \equiv \sigma/\mu$, where $\sigma$ is the standard deviation and $\mu$ is the mean. We estimate the coefficient of variation for all our median values of Fig.~\ref{DependenciaBin2}a \& \ref{DependenciaBin1}a, obtaining a typical $C_v=0.01$ for the metallicity median points, and $C_v=0.02$ for the \Ms\ median points. This means that the degree of variation around the median points of metallicity is of order 1$\%$, and 2$\%$ for the \Ms\ median points, allowing us to infer the physical dependencies through both representations.

\subsection[]{Some special cases: the Z-SFR and Z-SSFR relations}\label{ZSFRrelations}

Although the dependencies previously shown exhibit different median values depending  on the assumed binning direction, the general tendencies are similar (e.g., $Z$ and SFR always increase with  \Ms). However, the $Z$--{\it SFR} and $Z$--{\it SSFR} relations are special cases because they have a high scatter, and hence the median values  drastically change depending  on the binning direction taken, and care must be taken when interpreting the data.

Figure \ref{DependenciaBin2p1}e shows how the median metallicity varies in bins of SFR and \Ms. The median metallicity of galaxies increases as SFR increases for high mass galaxies, but  decreases as SFR increases for low mass galaxies. Following the same model described in $\S\,\ref{SFRonMZhorizontal}$, these changes can be attributed to different  gas content in galaxies at different stellar masses.

% Similarly to Fig. \ref{DependenciaBin2}a, we see a reverse at $\log(M/M_{\odot})\sim$10.2. For $\log(M/M_{\odot})\gtrsim$10.2, the metallicity increases with the SFR, For $\log(M/M_{\odot})\lesssim$10.2 however, the metallicity decreases when the SFR increases. 

On the other hand, Fig. \ref{DependenciaBin1p1}e shows the same relation as Fig. \ref{DependenciaBin2p1}e  but now considering the median SFR in bins of $Z$  and \Ms. The median SFR  values  indicate that SFR  increases when metallicity increases for massive galaxies. For low mass galaxies however, the SFR remains almost constant, or slightly decreasing when metallicity increases. Therefore, independenly of the binning direction taken, the result drive us to similar conclusions.

Figure \ref{DependenciaBin2p1}f shows the $Z$--{\it SSFR} relation taking the median $Z$ in bins of  SSFR and \Ms. In this case,  a reverse  is observed around log(SSFR) $\sim-$10. For log(SSFR) $\gtrsim-$10, the $Z$ of galaxies decreases when the SSFR increases, and vice-versa for log(SSFR) $\lesssim-10$. The same tendency is observed in the high redshift samples. This reverse is the same observed in Fig. \ref{DependenciaBin2}d, and agrees with the hypothesis of a different amount of gas for different stellar masses.
This reverse is more  evident  in Fig. \ref{DependenciaBin1p1}f,  that shows the same relationship but taking the median SSFR in bins of $Z$ and \Ms. Interestingly, the median SSFR values show two tails: one formed by low-mass galaxies displaying high SSFRs, in which the SSFR anticorrelates with $Z$ , and another formed by massive galaxies with low SSFRs, in which the SSFR correlates with $Z$ . Both tails converge to a log(SSFR) value of $\sim-$10.

This dual behaviour  is a consequence of the different physics involving low  and  high mass galaxies. The same explanation given in Sect \ref{SFRonMZhorizontal} is applicable here. A combination of   both downsizing, and the differences in the amount of neutral gas in galaxies can explain the bimodality  observed in those relationships. For low and high mass galaxies,  galaxies with a higher amount of gas  will show  higher SSFRs compared to galaxies at the same mass but with less HI.

Our explanation agrees with  \citet{Dave12}, who through an analytic formalism inspired by hydrodynamic simulations, describes an equilibrium between metallicity, gas fraction, and SFR. According to their predictions, at a given mass, galaxies that are gas rich and metal poor will have higher SFRs. On the other hand, gas poor  and metal rich galaxies will have  lower SFRs.

The high mass branch shown as pink points in Fig. \ref{DependenciaBin1p1}f shows a positive correlation between $Z$ and SSFR.  \citet{Lara13} find that the amount of massive galaxies with HI detected is too few to confirm the proposed physical explanation. It is also possible, however, that AGN feedback could be implicated in shutting down the SFR in massive galaxies. It is likely that the effectiveness of this process varies from one massive galaxy to another, probably depending on the history of its AGN activity. One could then imagine that the systems in which star formation was shut down earliest would have especially low SSFR and also low average metallicity, since there has been less opportunity for metals to be recycled into subsequent generations of stars. Galaxies which have experienced less efficient AGN feedback would be left with more gas, but will also have had more opportunity to recycle it, increasing their metallicity.

%***************************** ESTO NO LO HE TERMINADO.. PENDIENTE********************************
% If we suppose that massive galaxies have similar star formation histories, this will produce a correlation of SSFR with Z for high mass galaxies, and an anticorrelation for low mass galaxies, since this galaxies are more affected for baryonic material loss. A detailed view of this model will be given in Lara-L\'opez et al. (2012, in preparation)

% AQUI VOY!!!!
% NO OLVIDAR HABLAR DE LOS VIENTOS GALACTICOS PARA EXPLICAR QUE TIENEN MENOS METALES!!!

% Assuming that low mass galaxies will suffer a considerably loss of barionic material due to galacti winds, low mass galaxies with a high HI gas content will show higher SSFR and thus low $Z$ due to galactic winds.

% the SSFR and $Z$ are anticorrelated (see green  circles of Fig. \ref{DependenciaBin1p1}f). This implies that low mass

% show  a positive correlation between  SSFR and $Z$ (pink circles in Fig. \ref{DependenciaBin1p1}f)

% In this context, high mass galaxies with high HI gas are still processing their gas, and since their gravitational potencial due to its mass is high enough to retein their gas, they have reached higher metallicities than galaxies with less amount of gas  although all had similar stellar masses.

As shown in $\S\,\ref{SampleSelection}$ most of  the objects in our sample correspond to late--type galaxies. Nevertheless,  we studied the same relationships as a function of their S\'ersic index to explore another possible explanation  of this reverse as a function of morphology. However, we did not find any dependence with the S\'ersic index. Thus, this opposite behaviour between low- and high-mass galaxies seems likely to be explained primarily by the different amount of gas in each. A more detailed explanation of this model is given in \cite{Lara13}.

\begin{figure*}
\includegraphics[scale=0.87]{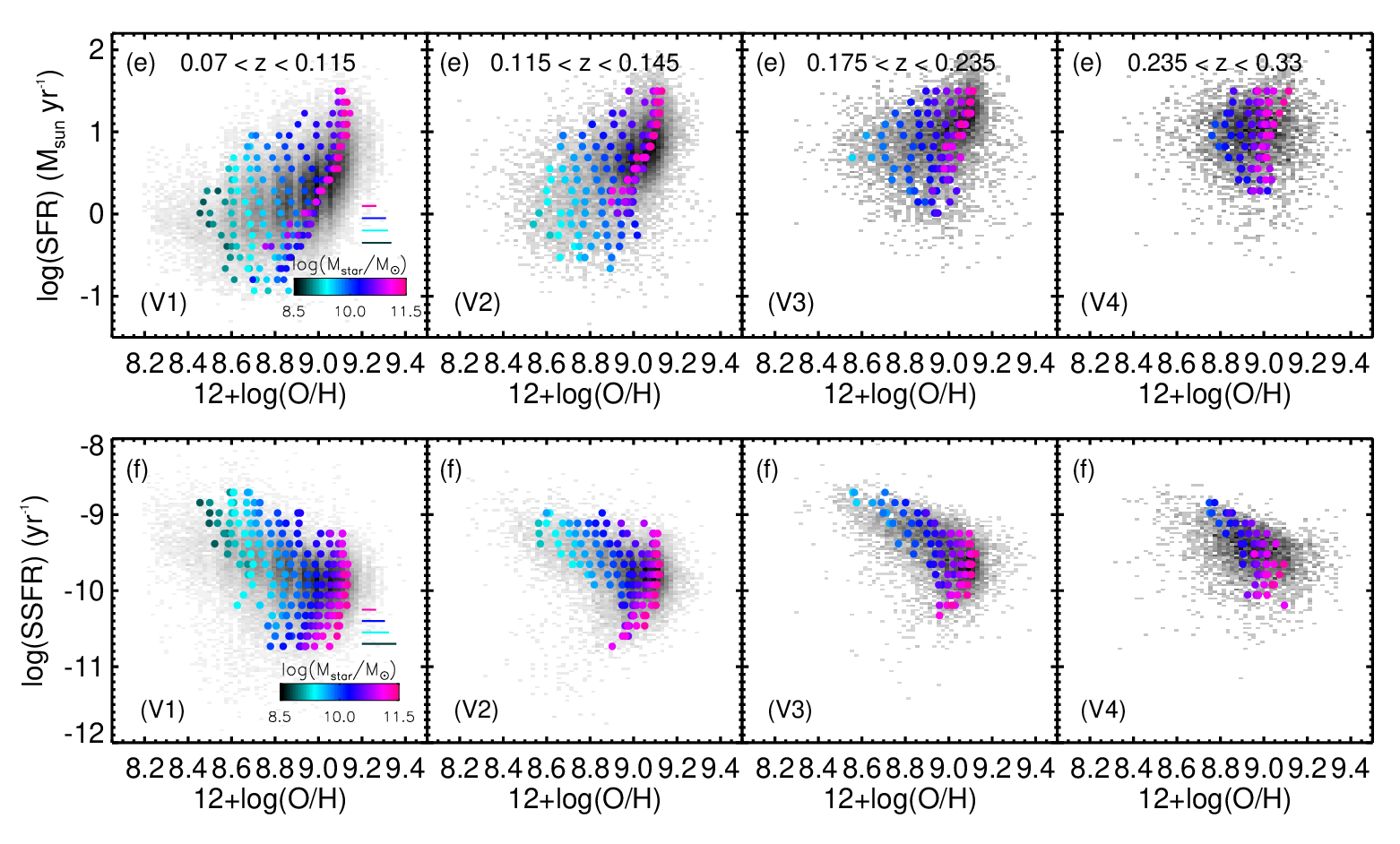}
\caption{From top to bottom, the \Ms\ dependence of the $Z$--{\it SFR} and $Z$--{\it SSFR} relations. 
The median metallicity in bins of SFR and SSFR for given bins of \Ms\ is shown in all cases.   The horizontal color lines show the 1-$\sigma$ dispersion of the horizontal variable of four representative bins, as shown in the color-bar of each panel.  The data density greyscale corresponds to the SDSS and GAMA sample together.}
\label{DependenciaBin2p1}
\end{figure*}

\begin{figure*}
\includegraphics[scale=0.87]{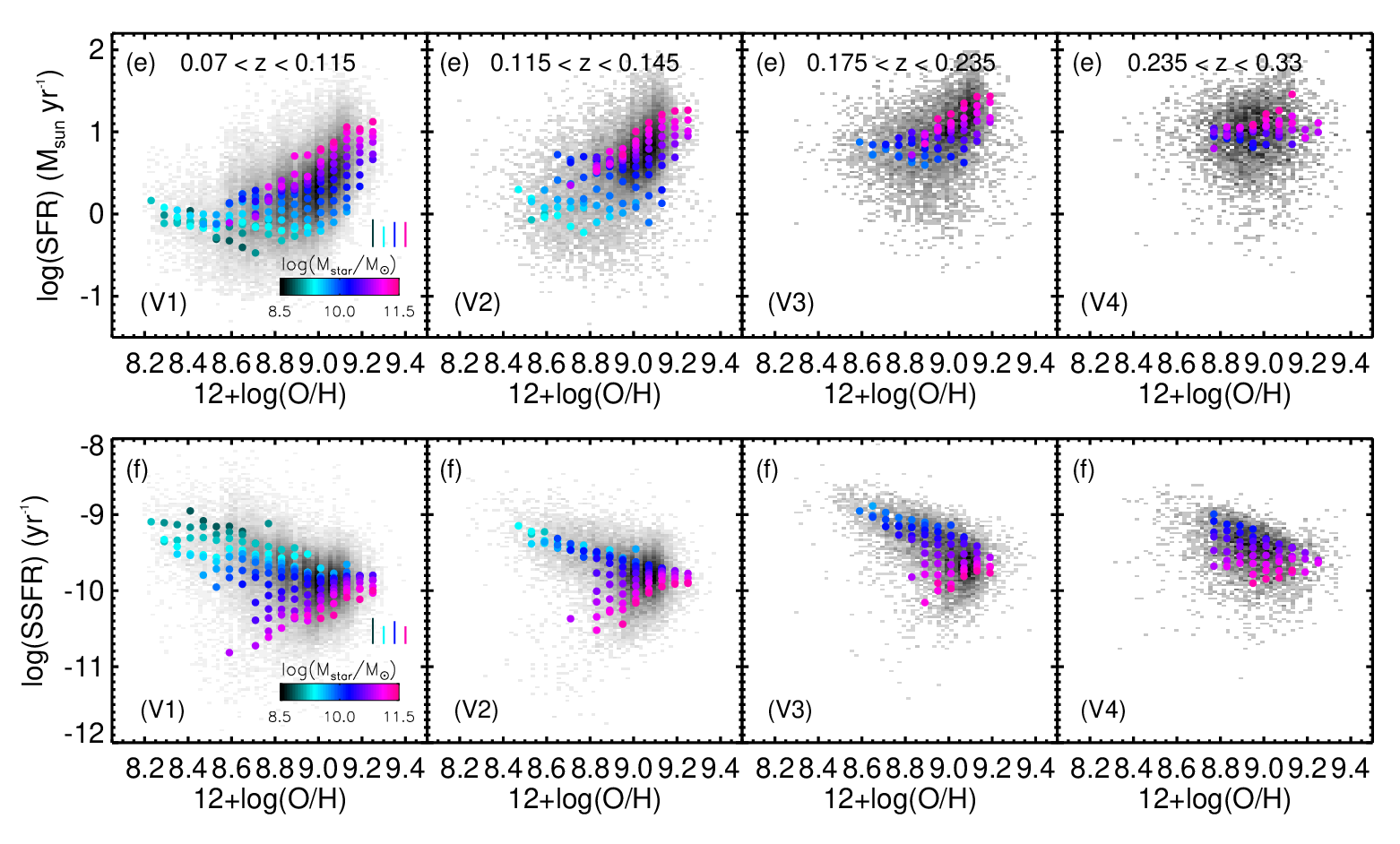}
\caption{From top to bottom, the \Ms\ dependence of the $Z$--{\it SFR} and $Z$--{\it SSFR} relation. 
The median SFR and SSFR in bins of metallicity for different bins of \Ms\ is plotted in all cases.  The vertical color lines show the 1-$\sigma$ dispersion of the vertical variable of four representative bins, as shown in the color-bar of each panel.  The gray density plot correspond to the SDSS and GAMA sample together.}
\label{DependenciaBin1p1}
\end{figure*}

\subsection[]{Dependencies using different metallicities and SFR indicators}\label{MetBias}

Here we analyze how strongly the \Ms$-Z$ relation depends on  several tracers of  SFR and metallicity. We use only our SDSS  volume-limited galaxy samples, and adopt three different methods to estimate metallicities:
\begin{enumerate}
\item the \citet{Tremonti04} Bayesian metallicities,
\item the {\NII}/{\OII} index with the calibration provided by \citet{Kewley02} and the update given by \citet{Kewley08}, and
\item the O3N2 index, defined in Eq. \ref{O3N2}, with the calibration of  \citet{Pettini04}. 
\end{enumerate}
For estimating the SFR we used the same two methods described in $\S\,\ref{SecGAMA}$,
\begin{enumerate}
\item the \citet{Brinchmann04} Bayesian SFRs, and 
\item the \citet{Hopkins03} estimations of the SFRs, which are based on the equivalent widths of \Ha.
\end{enumerate}

Figure \ref{DifMethodsZySFR} shows the SFR dependence on the \Ms$-Z$ relation  considering all the possible combinations of the  metallicity and SFR methods described above. Although  the dependencies vary somewhat from one combination to another, they all are observed in all these six combinations. It is noteworthy that the main change in the dependencies are due to the approach used in estimating the metallicities and not the SFRs.  This is consistent with recent work by, \citet{Zahid13} who find that the characteristic shape observed in Fig. \ref{DifMethodsZySFR} is related to the level of dust extinction.

\citet{Mannucci10} reported a different SFR dependency on the \Ms$-Z$ relation. 
 Following these authors,  all the SFR median points seem to converge at the high mass end. 
We attribute this behaviour to the use of the N2 method  \citep[which relies on the \NII~$\lambda$6584/\Ha\ ratio, see][]{Pettini04} for galaxies in the high-metallicity regime. It is well known that the N2 parameter is not useful  for 12+log(O/H)$>$ 8.8 \citep[e.g.][]{Yin07,LSD12}, as at high metallicities  the \NII~$\lambda$6584 saturates and it is no longer an appropriate tracer of the oxygen abundance. 
As a result of  using  a non-robust metallicity indicator, the high metallicity regime of \citet{Mannucci10} is saturated, and hence the dependency with SFR is lost,  showing an artificial convergence of SFR at high metallicites. See \citet{Lara12} for a detailed explanation of this issue.

\begin{center}
\begin{figure*}
\includegraphics[scale=0.8]{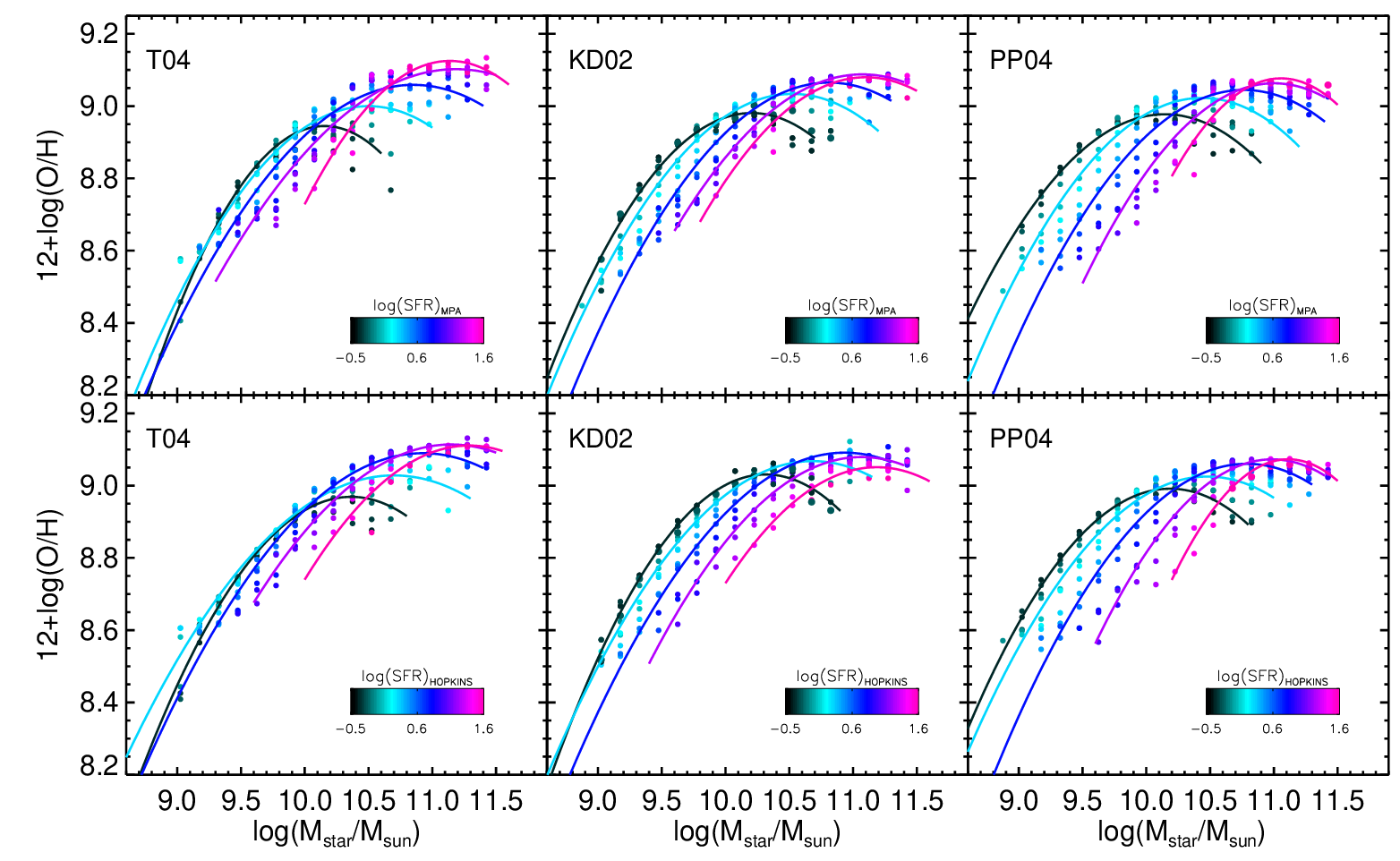}
\caption{\Ms$-Z$ relation for SDSS galaxies up to $z<0.365$. Left column shows the \Ms$-Z$ relation using the  \citet{Tremonti04} metallicities  and the \citet{Brinchmann04} and \citet{Hopkins03} SFRs. Middle and right column show the same relation but using the \citet{Kewley02}  and the \citet{Pettini04} metallicities, respectively, with the same SFRs described above.}
\label{DifMethodsZySFR}
\end{figure*}
\end{center}

\section{Summary and Conclusions}\label{Conclusions}

We studied the \Ms, $Z$, SFR, and SSFR dependencies and evolution, as well as the FP using SF galaxies of the GAMA and SDSS surveys.  We find a good agreement between the GAMA and SDSS survey in all cases.  Indeed, both surveys compliment  each other. We divided our whole sample in 16 volume-limited samples from redshift $z$=0.04 to $z$=0.365. Our findings can be summarised as follows:

\begin{itemize}

\item From a statistically robust sample combining the GAMA and SDSS surveys in volume limited samples up to z $<$ 0.1, we established a local \Ms$-Z$ relation. By fitting the zero point of the local \Ms$-Z$ relation, we quantified the metallicity evolution of every volume limited sample, finding a gradual decrement of metallicity as redshift increases, with a maximum evolution of $\sim$0.1 dex for the redshift range 0.330 $<z<$ 0.365. The evolution found is in agreement with the evolutionary models of \citet{Buat08}.

\item We studied the \Ms--{\it SFR} and \Ms--{\it SSFR} relationships as well. In a similar way as the \Ms$-Z$ relation, we established a local \Ms--{\it SFR} (--{\it SSFR}) relation, and fitted the zero point for all the volume limited samples at higher redshift. We found a maximum SFR evolution of $\sim$0.4 dex, and of $\sim$0.56 dex in SSFR.

\item We analysed the FP for the whole GAMA and SDSS galaxies, finding a good agreement between both samples and a single plane for both of them. The FP found allows us to recover the \Ms\ of SF galaxies through a  linear combination of  SFR and Z with a $\sigma$=0.2 dex.  The FP is a consequence of the direct dependence of the \Ms\ on the SFR and Z. The current mass in stars in a galaxy (\Ms) is a measure of the amount of gas currently being converted into stars (SFR), plus a measure of the star formation history, or past generations of stars, here represented by the metallicity (Z). 

\item No evidence of evolution has been shown in the FP in our sample (z $<$ 0.365), and there is a lack of evolution up to z$\sim$3.5 \citep[see][for details]{Lara10a}. This lack of evolution is a consequence of the SFR and Z evolving in different directions. While the SFR increases for high redshift galaxies, the Z decreases. Thus, when the linear combination to produce the FP is applied, those differences cancel out, allowing us to recover the \Ms\ of the galaxies.

\item We studied the dependencies and  relationships of the \Ms, $Z$, SFR, and SSFR, analysing all the possible combinations and binning directions. We studied the SFR and SSFR dependence on the \Ms$-Z$ relation, the Z dependence on the \Ms--{\it SFR} and \Ms--{\it SSFR} relations, and the \Ms\ dependence on the $Z-${\it SFR} and $Z-${\it SSFR} relationships. All of them showing a strong dependence and correlations to each variable.

\item We found that a correct interpretation is crucial when different binning directions are studied. For example, the SFR dependence in the \Ms$-Z$ relation apparently changes either if we estimate the median $Z$ in \Ms and SFR bins (Fig. \ref{DependenciaBin2}), or the median \Ms\ in Z and SFR bins (Fig. \ref{DependenciaBin1}). Nevertheless, the underlying physics is consistent with a single interpretation.

% AQUI HABLAR DE LA BIMODALIDAD

\item We found evidence of a reverse in the dependencies of low and high mass galaxies. For massive galaxies, the median metallicity is higher/lower for high/low SFR galaxies  at the same stellar mass. On the other hand, for low mass galaxies we find the opposite behaviour, the median metallicity is lower/higher for high/low SFR galaxies at the same stellar mass.

% \item We studied the peculiar case of the $Z$--{\it SFR}. In any of the other relationships studied,  the unveil nature of the estimated variable slightly changes the shape, but not the general tentency of the whole data. However, the \Ms\ dependencie on the $Z$--{\it SFR}  relation change the general tendency of the data depending on the binning direction taken. This relationship however, present a high scatter, implying that the contradictory tendencies can be just an artifact of the dispersion in the data.

\item  This reverse or bimodality is more evident in  $Z$--{\it SSFR}. Although this relation presents  a high scatter, we can see clearly two populations of galaxies, one formed by low mass galaxies showing an anticorrelation between SSFR and $Z$, and another population formed by massive galaxies showing a correlation between SSFR  and $Z$.

\item It is clear from all the dependencies that there is a different behaviour between low and high  mass galaxies. To explain this, we generated a  model based on the $Z$--{\it SSFR} relation. We propose that a combination of downsizing and different amount of neutral gas, can explain all our relationships and reverse observed.  According to this model, for a given stellar mass, and due to different amounts of neutral gas,  galaxies exhibit a wide range of SSFRs. The SSFR for the high mass galaxies correlates with metallicity (the SSFR increses when $Z$ increases). On the other hand, the SSFR of low mass galaxies anticorrelates with $Z$ (when the SSFR increases, the $Z$ decreases). This opposite behaviour at the low and high mass ends explains the reverse found in the \Ms$-Z$, \Ms--{\it SFR}, and \Ms--{\it SSFR} relationships.

\item We analysed the above dependencies using different combinations of SFR and metallicities, such as \citet{Hopkins03} and \citet{Brinchmann04} for the SFR, and \citet{Pettini04}. \citet{Tremonti04}, \citet{Kewley02} for the metallicity, In all the combinations, we found a strong dependence on the SFR for the \Ms$-Z$ relation, implying that this is independent of the method used.

\end{itemize}

\section*{Acknowledgments}

GAMA is a joint European-Australasian project based around a spectroscopic campaign using the Anglo-Australian Telescope. The GAMA input catalogue is based on data taken from the Sloan Digital Sky Survey and the UKIRT Infrared Deep Sky Survey. Complementary imaging of the GAMA regions is being obtained by a number of independent survey programs including GALEX MIS, VST KIDS, VISTA VIKING, WISE, Herschel-ATLAS, GMRT and ASKAP providing UV to radio coverage. GAMA is funded by the STFC (UK), the ARC (Australia), the AAO, and the participating institutions. The GAMA website is http://www.gama-survey.org/.
The work uses Sloan Digital Sky Survey (SDSS) data. Funding for the SDSS and SDSS-II was provided by the Alfred P. Sloan Foundation, the Participating Institutions, the National Science Foundation, the U.S. Department of Energy, the National Aeronautics and Space Administration, the Japanese Monbukagakusho, the Max Planck Society, and the Higher Education Funding Council for England. The SDSS was managed by the Astrophysical Research Consortium for the Participating Institutions. 
M. A. Lara-L\'opez thanks to the ARC for a super science fellowship, and to the $``$Summer School in Statistics for Astronomers$"$, Center for Astrostatistics, PennState, for invaluable tutorials on $``$R$"$ and PCA.

\clearpage

\appendix

\section[]{Metallicity and SFR calibrations}

We have calibrated the MPA-JHU metallicity  and SFR with other methods  using data from the SDSS. We have computed the PP04 metallicities for SDSS galaxies, and have calibrated them vs. the T04 metallicities using a linear fit, as shown in Fig. \ref{MetCalibration}. As a result, we obtained this calibration::

% 
% Garching vs. [NII/OII]
% 
% Para pasar del Metodo de KD02 al de Garching:
% 
% Met_KD02ToGarching= -0.70455967 +1.0813597*KD02
% 
% 
% Para pasar del PP04 al de Garching:
% 
%

\begin{equation}
% \rm 12+log(O/H)_{T04}=0.102655+1.021088 \; {\times} \; 12+log(O/H)_{PP04}
{\rm [  12+log(O/H) ] }_{\rm T04}=-1.0962 + 1.1570 \times {\rm [ 12+log(O/H) ]}_{\rm PP04}
\end{equation}

\begin{figure}
\includegraphics[scale=0.55]{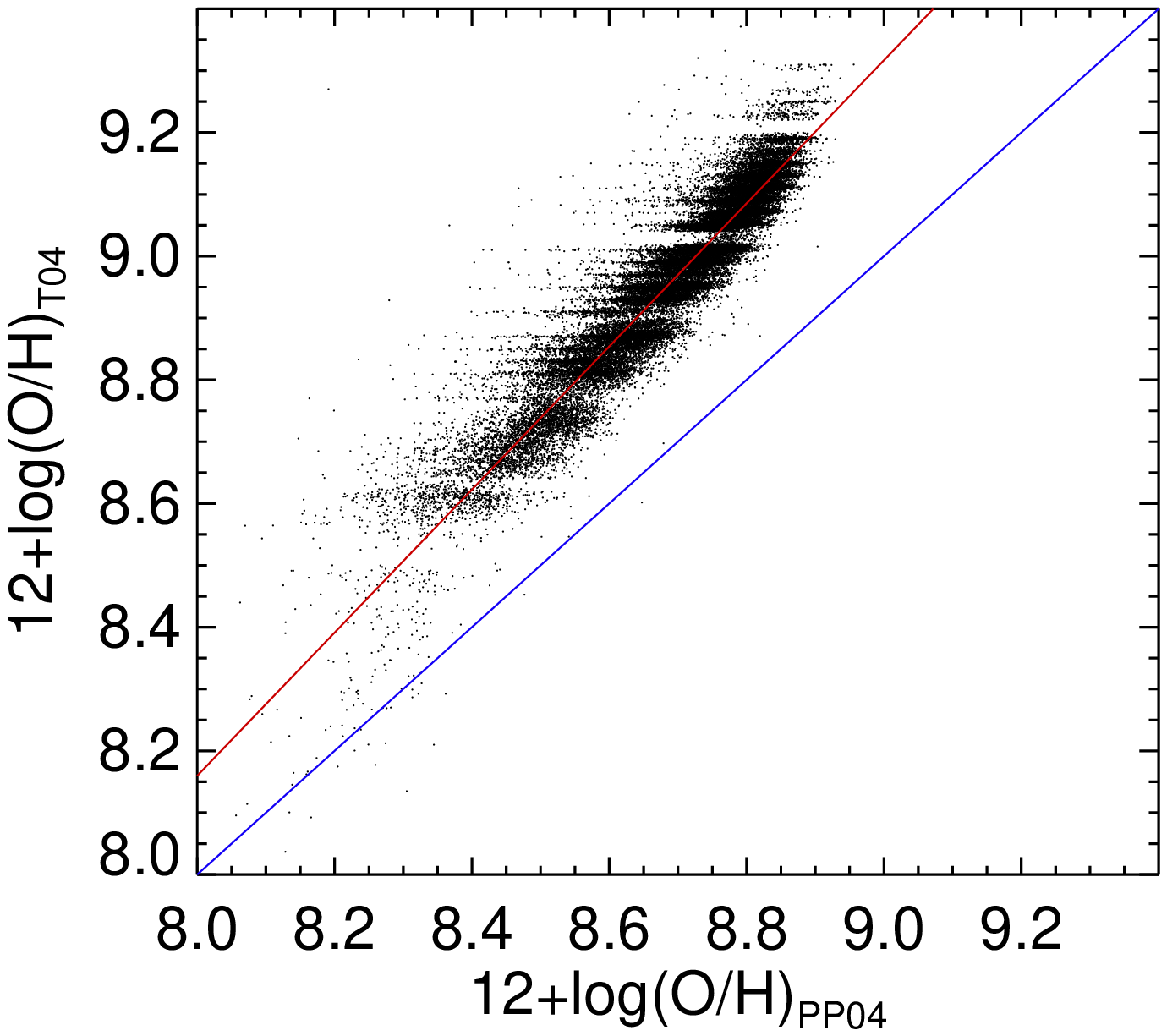}
\caption{Metallicity calibration from the Pettini $\&$ Pagel (2004) method compared to the  Tremonti et al. (2004) metallicities. The blue line shows the one to one relation, while the red line shows our calibration.}
\label{MetCalibration}
\end{figure}

% 
% Met_PP04ToGarching= 0.10226731+1.0210972*KD02
% 
% Alto z, muestras 8a 13
% 
% Met_PP04ToGarching=1.8460816+0.82528401*KD02
% 

We have also computed the Hopkins et al. (2004) SFRs for SDSS galaxies, and have calibrated them vs. the Brinchmann et al. (2004) metallicities using a linear fit, as shown in Fig. \ref{SFRCalibration}. As a result, we obtained this calibration::

\begin{equation}
\rm log(SFR)_{B04}=-0.064889 +1.37597 \;{\times}\; \rm log(SFR)_{Hop}
\end{equation} 
% 
% at higher redshift:
% 
% yMZSloan=-1.6306479 +0.79666592*xMZSloan

\begin{figure*}
\includegraphics[scale=0.55]{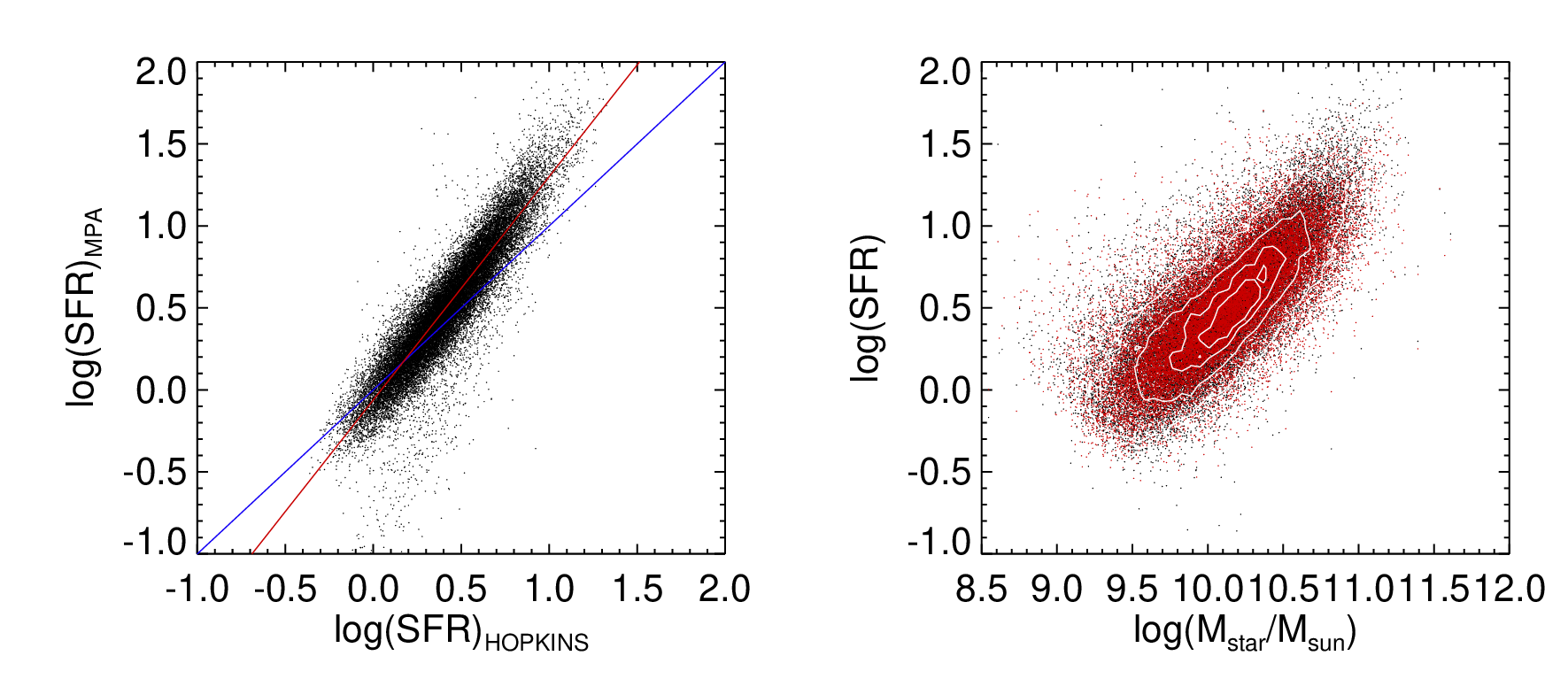}
\caption{Left: SFR calibration from the Hopkins et al. (2003) to the Brinchmann et al. (2004) SFRs. Blue line shows the one to one relation, while red line shows our calibration. Right: M-SFR relation using the  Brinchmann et al. (2004) SFRs (black), and the recalibrated SFRs (red).}
\label{SFRCalibration}
\end{figure*}

\section[]{GAMA and  SDSS counterparts, a metallicity comparison}

A metallicity comparison has been done matching the GAMA phase--I with the SDSS--DR7 catalogs. From both surveys, metallicities have been estimated for the counterparts exactly in the same way using the PP04. The result is plotted in Fig. \ref{CompPP04Sloan}. There is good agreement between GAMA and SDSS with a sigma of the residuals of $\sigma$=0.06.

\begin{figure}
\begin{center}
\includegraphics[scale=0.430]{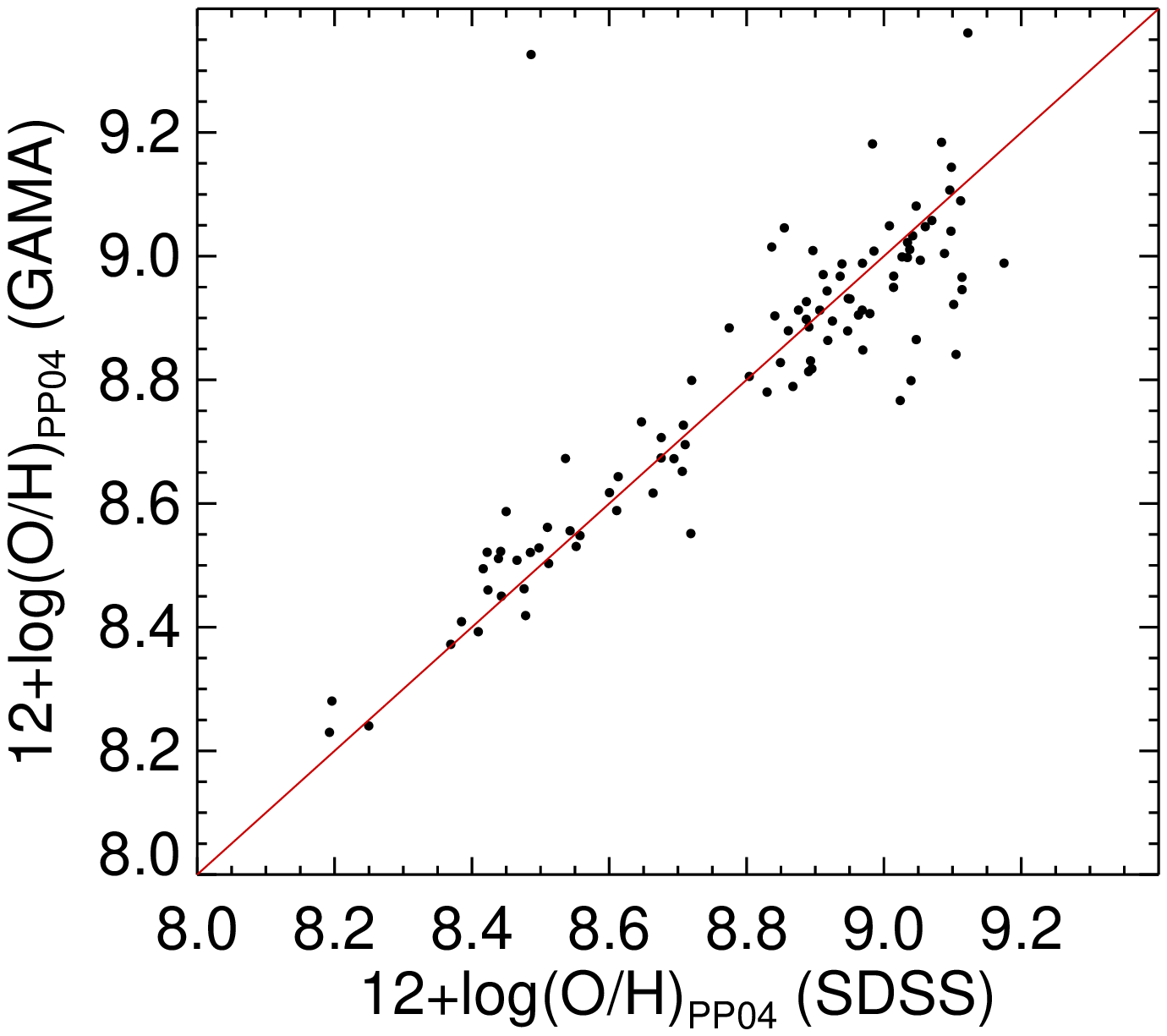}
\includegraphics[scale=0.430]{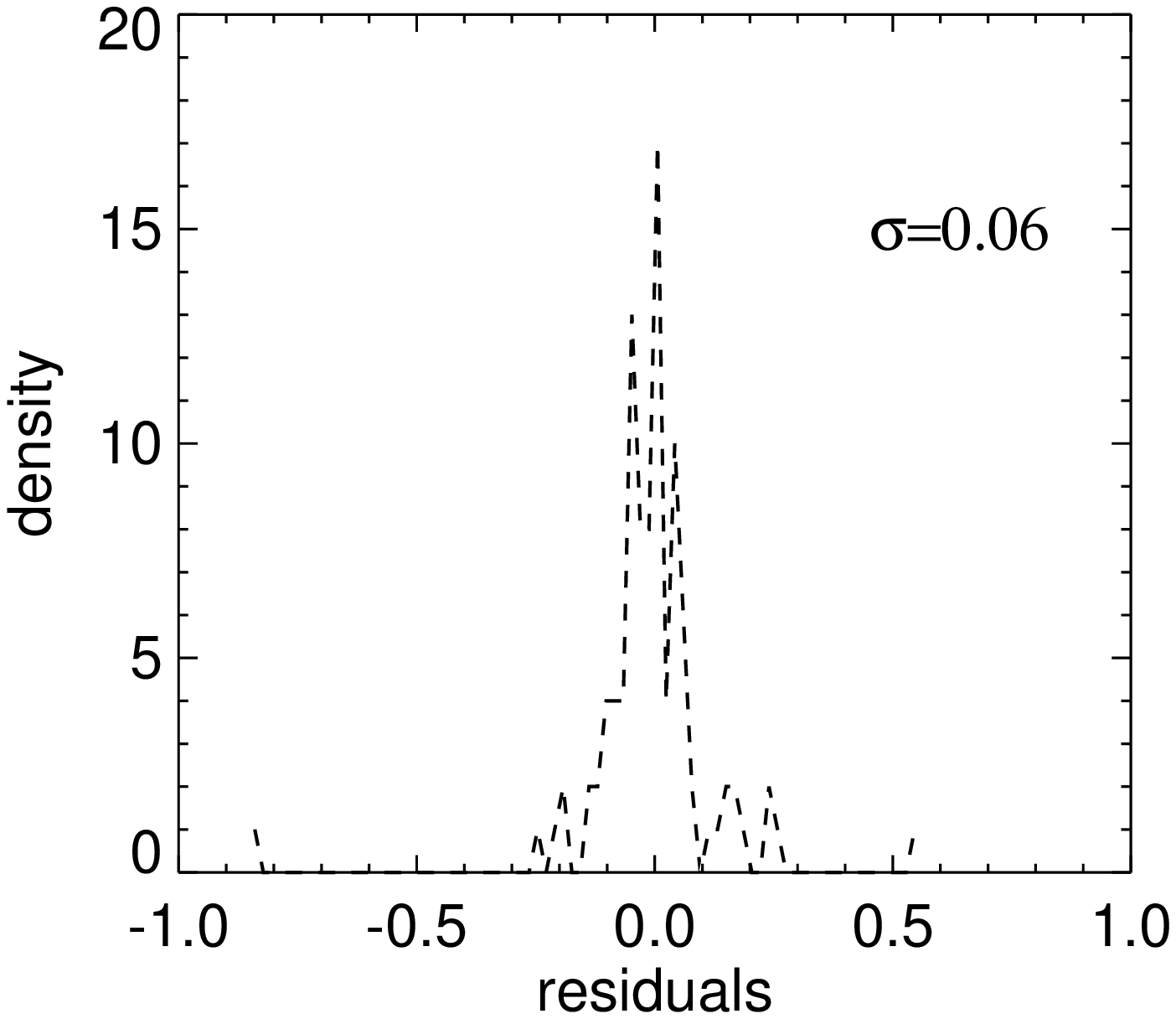}
\caption{Left: Comparison between GAMA and SDSS counterparts using the PP04 method, the red line corresponds to the one to one relation. Right: Histogram of the residuals.}
\label{CompPP04Sloan}
\end{center}
\end{figure}

\label{lastpage}

\end{document}